\documentclass[a4paper,fleqn,usenatbib,onecolumn]{mnras_arXiv}

\usepackage{mathptmx}

\usepackage[T1]{fontenc}
\usepackage{ae,aecompl}

\usepackage{graphicx}	
\usepackage{amsmath}	
\usepackage{amssymb}	
\usepackage{pdflscape}
\usepackage{multirow}
\usepackage{bm}

\numberwithin{equation}{section}

\title[Analytic models for the Rossiter-McLaughlin effect]{Analytic models
of the Rossiter-McLaughlin effect for arbitrary eclipser/star size
ratios and arbitrary multiline stellar spectra}

\author[R.V.~Baluev \& V.Sh.~Shaidulin]{
Roman V. Baluev$^{1,2}$\thanks{E-mail: r.baluev@spbu.ru} and Vakhit Sh. Shaidulin$^{2,1}$\\
$^1$Central Astronomical Observatory at Pulkovo of Russian Academy of Sciences, Pulkovskoje
shosse 65/1, St Petersburg 196140, Russia\\
$^2$Sobolev Astronomical Institute, St Petersburg State University, Universitetskij prospekt
28, Petrodvorets, St Petersburg 198504, Russia}

\date{Accepted 2015 September 28.
      Received 2015 September 26;
      in original form 2015 June 4}

\pubyear{2015}

\begin{document}
\label{firstpage}
\pagerange{\pageref{firstpage}--\pageref{lastpage}}
\maketitle

\begin{abstract}
We present an attempt to improve models of the Rossiter-McLaughlin effect by
relaxing several restrictive assumptions. We consider the entire multiline stellar spectrum
rather than just a single line, use no assumptions about the shape
of the lines profiles, and allow arbitrary size ratio for the star and its eclipser.
However, we neglect the effect of macro-turbulence and differential rotation. We construct
our model as a power series in the stellar rotation velocity, $V\sin i$, giving a closed
set of analytic formulae for up to three terms, and assuming quadratic limb-darkening law.
We consider three major approaches of determining the Doppler shift: cross-correlation with
a predefined template, cross-correlation with an
out-of-transit stellar spectrum, and parametric modelling of the spectrum.

A numerical testcase revels that our model preserves good accuracy for the
rotation velocity of up to the limit of $2-3$ times the average linewidth in the spectrum.
We also apply our approach to the Doppler data of HD~189733, for which we obtain
an improved model of the Rossiter-McLaughlin effect with two correction terms, and derive a
reduced value for $V\sin i$.
\end{abstract}




\section{Introduction}

Whereas the number of the discovered exoplanets grows continuously, the importance of their
cross-characterization by independent observation techniques
increases. There are two mostly productive planet detection methods: by radial
velocity (RV) variations and by a photometric fadening during a transit. Consequently, the
joint analysis of the combined RV+transit data gained a special value in the recent years.

This task is not reduced to a mere combination of the RV and transit data,
with their respective separate models. In such cases we may also observe
hybrid events, like the Rossiter-McLaughlin (RM) effect, which is basically a spectrosopic
view of a planetary transit before a rotating star.

The most simple model of this effect is based on the assumption that the
measured Doppler anomaly is equal to the average RV of the occulted stellar disk
\citep{Kopal42,Ohta05,Gimenez06}. We will call this as classic RM model.
This average velocity is given by an exact formula
\begin{equation}
V_\mathrm{mean} = -\frac{f v_{\mathrm p}}{1-f},
\label{vmean}
\end{equation}
where $f$ is a fraction of the flux blocked by the planet, and $v_{\mathrm p}$
is the average ``subplanet'' RV, computed with an account for the stellar limb
darkening. The remaining problem here is to compute $f$ and $v_{\mathrm p}$ for an assumed
limb-darkening law. However, $V_\mathrm{mean}$ in this formula is not the same
physical quantity as the Doppler anomaly that we seek. Instead of averaging the RV over the
stellar disk, we must average the stellar spectrum first, and then determine the Doppler
anomaly from this average spectrum.

The average of a function is not equal to the function of an averaged argument, unless the
function is linear or well-linearizable. Therefore, the formulae~(\ref{vmean}) approximates
the RM anomaly only if the stellar rotation velocity $V\sin i$ is very small. Ideally,
it should be much smaller than the typical width of the spectral lines (scaled
in the velocity units). In this case, stellar spectrum can be linearized with respect to
the rotational Doppler shift. In the remaining cases, the formula~(\ref{vmean}) cannot
be used to predict the RM anomaly, with enough accuracy at least.

There are works in which an attempt is made to construct more accurate approximation
than~(\ref{vmean}), see e.g. \citep{Hirano10,Hirano11,Boue13}. They succeeded a lot in this
field, but the problem is still far from being solved due to some restrictive asumptions
adopted by these authors. A major one is that their analytic results refer to a simplified
single-line model of the stellar spectra. In practice, however, Doppler shift is determined
from rich spectra that contain thousands of lines or more. Assumption of a single line
cannot leave no implications on the reliability of the model. Another important assumptions
are that line profiles should be symmetric \citep{Boue13} and the planet
is assumed small in all these works.

Our aim here is to consider the full stellar spectrum containing multiple lines, and
discuss the differences with a single-line model. Also, we tried to avoid decompositions in
planet radius, whenever possible. This may be useful for red dwarfs transited by a giant
planet. In this case, the planet/star radii ratio may exceed $1/10$. As far as
we could learn, the largest or one of the largest values for
this ratio currently belongs to the unique circumbinary planet KIC~9632895
\citep{Welsh14}. Here this ratio reaches $0.26$ for one of the binary components, although
the absolute planet radius is only $6.2 R_{\earth}$. In theory, a red dwarf star can
be even smaller than a giant planet, so such a ratio can even be comparable to or exceed
unit.

The structure of the paper is summarized as follows. In Sect.~\ref{sec_anom} we give
general mathematical formulation and present several methods and main formulae
that are useful for analytic modelling the RM effect under different assumptions. In
Sect.~\ref{sec_rmobs} we derive our main results of the RM effect.
In Sect.~\ref{sec_mom} we describe an analytic computation of the RV momenta in an occulted
stellar disk, which appear in our RM model. In Sect.~\ref{sec_num} we present results of a
simulation to test the accuracy and usefullness of the model. In Sect.~\ref{sec_appl} we
apply our models of the RM effect to the public data of the MS star HD~189733, using it as
a testcase.

\section{Main mathematical methods and techniques for modelling the Rossiter-McLaughlin anomaly}
\label{sec_anom}

\subsection{General formulae and definitions}

Let us adopt the logarithmic scale in the wavelength, $s=\ln\lambda$, and denote the
spectrum of the star surface near the disk centre as $\mathcal
F(s)$. The Doppler-shifted spectrum should then be $\mathcal F(s-\tau)$, where the Doppler
shift is $\tau=v_z/c+\mathcal O(v^2/c^2)$, with $v_z$ being the radial velocity of
an emitting point (the $z$ axis directed along the line of sight). This is a
non-relativistic approximation. Light coming from different points
in the visible stellar disk is combined with different Doppler
shifts and different local brightness, forming two auxiliary spectra: the cumulative
star spectrum $\mathcal F_\star(s)$ and the ``subplanet'' spectrum $\mathcal
F_{\mathrm p}(s)$, which is generated by a portion of the surface blocked by the transiting
object. These spectra can be expressed as follows:
\begin{equation}
\mathcal F_\star(s) = \int\limits_{|\mathbfit R|<1} \mathcal F(s-\upsilon x) I(|\mathbfit R|,s) d\mathbfit R, \qquad
\mathcal F_{\mathrm p}(s) = \int\limits_{\mathcal S_{\mathrm p}} \mathcal F(s-\upsilon x) I(|\mathbfit R|,s) d\mathbfit R, \qquad
\qquad \mathbfit R=\{x,y\}.
\label{spectra}
\end{equation}
where $\upsilon=\frac{V\sin i}{c}$ is a renormalized rotation velocity, $I(R,s)$
is the limb-darkening law normalized to $I(0)=1$. This law may depend on the wavelength.
The integration is done either over the entire star disk $|\mathbfit R|<1$ or
over the subplanet portion of the disk $\mathcal S_\mathrm{p}$. The star radius is assumed
unit here, meaning that radial velocity of each point of the surface is equal to
just $\upsilon x$. The observed star spectrum during a transit is then expressed
as $\mathcal F_\mathrm{t}(s)=\mathcal F_\star(s)-\mathcal
F_{\mathrm p}(s)$. The formulae~(\ref{spectra}) assume that their integrad does not depend
on the point in the stellar disk, except for via the rotational Doppler shift and limb
darkening law that may vary with wavelength. Some effects may induce additional changes.
For example, macro-turbulence in the stellar
atmosphere makes lines characteristics different in the disk centre and near the
limb due to different projected geometry of the turbulent motions
\citep[e.g.][]{Hirano11,Boue13}. Here we do not take into account effects of this type.

From~(\ref{spectra}), the spectrum of a non-rotating star would be
\begin{equation}
\mathcal F_\star^0(s) = \mathcal F(s) \int\limits_{|\mathbfit R|<1} I(|\mathbfit R|,s) d\mathbfit R,
\label{nonrot}
\end{equation}
which slightly differs from the surface spectrum $\mathcal F(s)$ due to the wavelength
dependence of the limb-darkening law. However, the multiplier near $\mathcal F$ is a slowly
varying function, so in practice the difference between $\mathcal F_\star^0$ and $\mathcal
F$ is not important. Below, we will often say ``non-rotating star spectrum''
actually meaning $\mathcal F$.

Contrary to \citet{Boue13}, we do not make an assumption that $\mathcal F(s)$ contains only
a single line, and also we honor the dependence of the limb-darkening law on the
spectral range. Concerning the notations, we do not introduce an explicit Doppler shift to
the argument of $\mathcal F_{\mathrm p}$ at this stage, and we do not normalize our spectra
to unit.

\subsection{Modelling the procedure of determining the Doppler shift from the spectrum}

Now assume that we have a comparison, or template, spectrum $\mathcal F_\mathrm{T}(s)$ and
seek the best fitting Doppler shift by minimizing the goodness-of-fit function as follows:
\begin{equation}
\chi^2(\hat s, a) = \int\limits_{-\infty}^{+\infty} \left(\mathcal F_\mathrm{t}(s) - a\mathcal F_\mathrm{T}(s-\hat s) \right)^2 ds \longmapsto \min_{\hat s, a}.
\label{chi2b}
\end{equation}
From now on, let us introduce the scalar product of functions $\langle f,g \rangle$ and the
norm $||f||^2$ in the sense of the $L_2$ metric. With these definitions we may write the
following:
\begin{equation}
\chi^2(\hat s, a) = ||\mathcal F_\mathrm{t}||^2 - 2a \left\langle \mathcal F_\mathrm{t}(s)\mathcal F_\mathrm{T}(s-\hat s) \right\rangle + a^2 || \mathcal F_\mathrm{T}||^2.
\label{chi2}
\end{equation}
The first and the third terms here do not depend on $\hat s$, so to fit $\hat s$ means
to maximize the cross-correlation function (CCF):
\begin{equation}
C_\mathrm{tT}(\hat s) = \left\langle \mathcal F_\mathrm{t}(s)\mathcal F_\mathrm{T}(s-\hat s) \right\rangle \longmapsto \max_{\hat s}
\quad \implies \quad
\left\langle \mathcal F_\mathrm{t}(s)\mathcal F_\mathrm{T}'(s-\hat s) \right\rangle=0.
\label{ccf}
\end{equation}
Note that without loss of generality we may assume that CCF of $\mathcal F_\star$ with
$\mathcal F_\mathrm{T}$ is maximized at $\hat s=0$, implying that
\begin{equation}
\left\langle \mathcal F_\star\mathcal F_\mathrm{T}' \right\rangle = 0.
\label{ccfstar}
\end{equation}
This means that the template $\mathcal F_\mathrm{T}$ is centred so that for an
uneclipsed star the fitted RV is zero, and during the transit we deal with only the RV
offset due to the RM effect.\footnote{In either case, we do not take
into account the Doppler shift due to the motion of the star around the star-planet system
barycentre. We always consider only Doppler shifts relatively to the star orbital motion.}

The comparison template $\mathcal F_\mathrm{T}$ may be either an a
priori given mask, or we may adopt it be equal to the uneclipsed star spectrum $\mathcal
F_\mathrm{T} = \mathcal F_\star$. Up to a certain degree, these cases model the classic CCF
approach, and to the iodine cell technique of Doppler measurements, respectively
\citep{Hirano10,Hirano11,Boue13}. Note that \citet{Boue13} say that the line profile should
be symmetric to have~(\ref{chi2b}) be equivalent to~(\ref{ccf}). We believe this
requirement is excessive, because their integral $I_2$ in their eqs.~(18,19) is always
zero, even when the profile is asymmetric. This can be established by integrating it
by parts, taking into account that boundary effects are negligible if the total integration
range is large (they actually assume it is infinite). In fact, this integral is equal to
the derivative of our $||\mathcal F_\mathrm{T}||^2$ over $\hat s$, but this norm does not
depend on any shift in the integration variable (again, neglecting the boundary effects).

Of course, the practical procedures of determining $\hat s$ are always
more complicated than in the approximations adopted above. For example, when dealing with a
predefined template mask $\mathcal F_\mathrm{T}$, the CCF $C_{\mathrm{tT}}(s)$ is actually
not directly maximized but first fitted by a
Gaussian $\mathcal G_{\sigma}(s-\hat s)$ via $\hat s$ and $\sigma$, and then the
fitted value of $\hat s$ is adopted as a Doppler shift estimate
\citep{Baranne96,Pepe02}. In this case we should solve a secondary $\chi^2$-minimization
task:
\begin{equation}
\chi^2_\mathrm{CCF}(\hat s, \sigma, a) = \int\limits_{-\infty}^{+\infty} \left( C_\mathrm{tT}(s) - a\mathcal G_{\sigma}(s-\hat s) \right)^2 ds = ||C_\mathrm{tT}||^2 - 2a \left\langle C_\mathrm{tT}(s) \mathcal G_{\sigma}(s-\hat s)\right\rangle + a^2 ||\mathcal G_{\sigma}||^2 \longmapsto \min_{\hat s, \sigma, a}.
\label{chi2b2}
\end{equation}
Obviously, finding $\hat s$ is again equivalent to maximizing just a
CCF, but already of a second-level one:
\begin{equation}
\tilde C_\mathrm{tT}(\hat s | \sigma) = \left\langle C_\mathrm{tT}(s) \mathcal G_{\sigma}(s-\hat s) \right\rangle =
\iint\limits_{-\infty}^{+\infty} \mathcal F_\mathrm{t}(u)\mathcal F_\mathrm{T}(u-s) \mathcal G_{\sigma}(s-\hat s)\, du\, ds =
\left\langle \mathcal F_\mathrm{t}(s) \tilde{\mathcal F}_\mathrm{T}(s-\hat s | \sigma)\right\rangle, \quad
\tilde{\mathcal F}_\mathrm{T}(s|\sigma) = \int\limits_{-\infty}^{+\infty} \mathcal F_\mathrm{T}(s-u) \mathcal G_{\sigma}(u)\, du.
\label{ccf2}
\end{equation}
As we can see, this method becomes equivalent to the one with direct CCF maximization,
if we replace the original template $\mathcal F_\mathrm{T}$ by $\mathcal F_\mathrm{T}$
convolved with the fitted Gaussian (thus imposing some broadening effect on the lines of
$\mathcal F_\mathrm{T}$). But now it becomes important that $\tilde{\mathcal F}_\mathrm{T}$
depends on the parameter $\sigma$, which should be fitted simultaneously with $\hat s$. The
best fitting values of $a$ and $\sigma$ can be obtained by equating the partial derivatives
of~(\ref{chi2b2}) to zero. Taking into account the transformation~(\ref{ccf2})
we finally obtain an implicit equation for $\sigma$:
\begin{equation}
\left\langle \mathcal F_\mathrm{t}(s) \tilde{\mathcal F}_\mathrm{T}(s-\hat s|\sigma)\right\rangle = \frac{a}{2\sigma\sqrt\pi}, \quad
\left\langle \mathcal F_\mathrm{t}(s) \tilde{\mathcal F}''_\mathrm{T}(s-\hat s|\sigma)\right\rangle = -\frac{a}{4\sigma^3\sqrt\pi} \qquad \implies \quad
2\sigma^2 = - \frac{\left\langle \mathcal F_\mathrm{t}(s) \tilde{\mathcal F}_\mathrm{T}(s-\hat s|\sigma)\right\rangle}{\left\langle \mathcal F_\mathrm{t}(s) \tilde{\mathcal F}''_\mathrm{T}(s-\hat s|\sigma)\right\rangle}.
\label{sigma}
\end{equation}
Here we used an identity $\partial\mathcal G_\sigma/\partial\sigma = \sigma
\partial^2\mathcal G_\sigma/\partial s^2$, implying that
$\partial\tilde{\mathcal F}_\mathrm{T}/\partial\sigma = \sigma \tilde{\mathcal
F}_\mathrm{T}''$. By a convention, the stroke always refers to derivatives with respect to
$s$, not $\sigma$. Additionally, instead of~(\ref{ccfstar}), we must satisfy analogous
equation for $\tilde{\mathcal F}_\mathrm{T}$:
\begin{equation}
\left\langle \mathcal F_\star(s) \tilde{\mathcal F}_\mathrm{T}'(s|\sigma_0) \right\rangle = 0, \qquad
2\sigma_0^2 = - \frac{\left\langle \mathcal F_\star(s) \tilde{\mathcal F}_\mathrm{T}(s|\sigma_0)\right\rangle}{\left\langle \mathcal F_\star(s) \tilde{\mathcal F}''_\mathrm{T}(s|\sigma_0)\right\rangle}
\label{ccfstar2}
\end{equation}

The iodine cell techniques \citep{Butler96,AngladaEscudeButler12} are also much more
complicated then the simplified fitting like~(\ref{chi2b}). In particular, the number of
spectral parameters is much larger than two. In this work we adopt~(\ref{chi2b})
with $\mathcal F_\mathrm{T} = \mathcal F_\star$ as an approximation to the reality. In this
approximation, the resulting Doppler shift should become the same as if we plainly
maximized the CCF with $\mathcal F_\star$. The method of cross-correlating with a reference
star spectrum $\mathcal F_\star$ is, by the way, another independent Doppler technique that
is used in practice sometimes \citep{Lanotte14}.

\subsection{Two types of approximations leading to a ``small'' RM anomaly}

To move any further from~(\ref{ccf}), we may need to assume that $\hat s$ is small
enough to justify the power series decomposition in $\hat s$. This assumption becomes valid
when one of the following is satisfied:
\begin{enumerate}
\item Rotation velocity $V\sin i$ is small enough in comparison with typical line widths
(in the spectrum of a non-rotating star). In this case we can decompose both spectra
$\mathcal F_\star$ and $\mathcal F_{\mathrm p}$ into powers of $\upsilon$. Regardless of
this restriction, the size of the transiting object can be arbitrary here, e.g. comparable
to the star itself or even larger. Also, this approach does not need to make assumptions
about shapes of spectral lines.

\item Relative flux drop $f$ during the transit is small enough, so that $\mathcal
F_{\mathrm p}$ causes only a small anomaly to each line in the combined spectrum $\mathcal
F_\mathrm{t}$. However, this anomaly may be shifted significantly, even by a
quantity larger or much larger than typical line widths for a non-rotating star. In this
case we can introduce various power-series decompositions in $f$, but we
cannot decompose $\mathcal F_\star$, and hence $\mathcal F_\mathrm{t}$. So, we have to
either use more or less realistic approximations of the line shapes (e.g. assume
they are Gaussian) or to use numeric computations where required. This is the approach
adopted by \citet{Boue13}. Note that in this method it is still legal to decompose
$\mathcal F_{\mathrm p}$ into powers of Doppler shift after a proper centering, because if
the planet is small it blocks only a small range of surface rotation velocities, well below
the typical line widths.
\end{enumerate}
Note that e.g. \citet{Hirano10} uses both these assumptions simultaneously.

Regardless of which of the above assumptions is adopted, let us first handle the necessary
decomposition of $\mathcal F_\mathrm{T}'$ in~(\ref{ccf}):
\begin{equation}
\left\langle \mathcal F_\mathrm{t}\mathcal F_\mathrm{T}' \right\rangle -
\hat s \left\langle \mathcal F_\mathrm{t}\mathcal F_\mathrm{T}'' \right\rangle +
\frac{\hat s^2}{2} \left\langle \mathcal F_\mathrm{t}\mathcal F_\mathrm{T}''' \right\rangle -
\frac{\hat s^3}{6} \left\langle \mathcal F_\mathrm{t}\mathcal F_\mathrm{T}'''' \right\rangle +
\mathcal O({\hat s}^4)= 0.
\label{RVeq}
\end{equation}
Note that by using~(\ref{ccfstar}) we may derive that $\langle \mathcal F_{\mathrm
t}\mathcal F_\mathrm{T}' \rangle = - \langle \mathcal F_{\mathrm p}\mathcal F_\mathrm{T}'
\rangle$. Also, we may perform an integration by parts in any scalar product
of the type $\langle F^{(k)} G^{(m)} \rangle$ to move differentiations from one its operand
to another, when necessary.

The solution for $\hat s$ can be derived from~(\ref{RVeq}) by successive approximations,
and the first three terms look like:
\begin{equation}
\hat s \simeq - \frac{\langle \mathcal F_{\mathrm p}\mathcal F_\mathrm{T}' \rangle}{\langle \mathcal F_\mathrm{t}\mathcal F_\mathrm{T}'' \rangle} +
\frac{1}{2} \left(\frac{\langle \mathcal F_{\mathrm p}\mathcal F_\mathrm{T}' \rangle}{\langle \mathcal F_\mathrm{t}\mathcal F_\mathrm{T}'' \rangle}\right)^2 \frac{\langle \mathcal F_\mathrm{t}\mathcal F_\mathrm{T}''' \rangle}{\langle \mathcal F_\mathrm{t}\mathcal F_\mathrm{T}'' \rangle} -
\frac{1}{2} \left(\frac{\langle \mathcal F_{\mathrm p}\mathcal F_\mathrm{T}' \rangle}{\langle \mathcal F_\mathrm{t}\mathcal F_\mathrm{T}'' \rangle}\right)^3 \left(\frac{\langle \mathcal F_\mathrm{t}\mathcal F_\mathrm{T}''' \rangle}{\langle \mathcal F_\mathrm{t}\mathcal F_\mathrm{T}'' \rangle}\right)^2 +
\frac{1}{6} \left(\frac{\langle \mathcal F_{\mathrm p}\mathcal F_\mathrm{T}' \rangle}{\langle \mathcal F_\mathrm{t}\mathcal F_\mathrm{T}'' \rangle}\right)^3 \frac{\langle \mathcal F_\mathrm{t}\mathcal F_\mathrm{T}'''' \rangle}{\langle \mathcal F_\mathrm{t}\mathcal F_\mathrm{T}'' \rangle}
\label{fitRV}
\end{equation}
The first-order approximation $\hat s_1 = - \langle \mathcal F_{\mathrm p}\mathcal
F_\mathrm{T}' \rangle/ \langle \mathcal F_\mathrm{t}\mathcal F_\mathrm{T}'' \rangle$ is
a small quanitity, so~(\ref{fitRV}) represents actually a power series
in $\hat s_1$. Its error is then $\mathcal O(\hat s_1^4)$.

If the Doppler shift is determined by fitting the CCF with a Gaussian,
as in~({\ref{chi2b2}}), we should replace $\mathcal F_\mathrm{T}$ with
$\tilde{\mathcal F}_\mathrm{T}$ and also need to provide an approximation for two variables
$\hat s, \sigma$. To reach this goal, we consider the system of two equations, $\tilde
C_\mathrm{tT}'(\hat s)=0$ and the last one
in~(\ref{sigma}) for $\sigma$, and linearize them about the point $\hat s=0$ and
$\sigma=\sigma_0$. Taking into account~(\ref{ccfstar2}),
this yielded the following first-order approximation:
\begin{equation}
\frac{\sigma^2 - \sigma_0^2}{2} \simeq \frac{\left\langle \mathcal F_\mathrm{p} \tilde{\mathcal F}_\mathrm{T}' \right\rangle \left\langle \mathcal F_\mathrm{t} \tilde{\mathcal F}_\mathrm{T}''' \right\rangle + \left\langle \mathcal F_\mathrm{t} \tilde{\mathcal F}_\mathrm{T}'' \right\rangle \left(\left\langle \mathcal F_\mathrm{p} \tilde{\mathcal F}_\mathrm{T} \right\rangle + 2\sigma_0^2 \left\langle \mathcal F_\mathrm{p} \tilde{\mathcal F}_\mathrm{T}'' \right\rangle \right)}{5\left\langle \mathcal F_\mathrm{t} \tilde{\mathcal F}_\mathrm{T}'' \right\rangle \left\langle \mathcal F_\mathrm{t} \tilde{\mathcal F}_\mathrm{T}'''\right\rangle - 2\sigma_0^2 \left(\left\langle \mathcal F_\mathrm{t} \tilde{\mathcal F}_\mathrm{T}'''\right\rangle^2 - \left\langle \mathcal F_\mathrm{t} \tilde{\mathcal F}_\mathrm{T}''\right\rangle \left\langle \mathcal F_\mathrm{t} \tilde{\mathcal F}_\mathrm{T}'''' \right\rangle \right)}, \quad
\hat s \simeq \frac{- \left\langle \mathcal F_\mathrm{p} \tilde{\mathcal F}_\mathrm{T}' \right\rangle + \left\langle \mathcal F_\mathrm{t} \tilde{\mathcal F}_\mathrm{T}''' \right\rangle \frac{\sigma^2-\sigma_0^2}{2}}{\left\langle \mathcal F_\mathrm{t} \tilde{\mathcal F}_\mathrm{T}''\right\rangle}.
\label{fitRV2}
\end{equation}
For shortness, $\tilde{\mathcal F}_\mathrm{T}$ without arguments corresponds
to $\sigma=\sigma_0$ here. In what follows below, we do not need more terms in the
decomposition~(\ref{fitRV2}).

\subsection{Comparison with \citep{Boue13}}

\citet{Boue13} assume that $\mathcal F_{\mathrm p}\propto f$ and use only the first-order
approximation in $f$. In this case our formula~(\ref{fitRV}) can be reduced as follows
\begin{equation}
\hat s \simeq - \frac{\left\langle \mathcal F_{\mathrm p}\mathcal F_\mathrm{T}' \right\rangle}{\left\langle (\mathcal F_\star-\mathcal F_{\mathrm p}) \mathcal F_\mathrm{T}'' \right\rangle} \simeq
 - \frac{\left\langle \mathcal F_{\mathrm p}\mathcal F_\mathrm{T}' \right\rangle}{\left\langle \mathcal F_\star\mathcal F_\mathrm{T}'' \right\rangle},
\label{fitRVf}
\end{equation}
and formulae~(\ref{fitRV2}) turns in the similar way into
\begin{equation}
\frac{\sigma^2 - \sigma_0^2}{2} \simeq \frac{\left\langle \mathcal F_\mathrm{p} \tilde{\mathcal F}_\mathrm{T}' \right\rangle \left\langle \mathcal F_\star \tilde{\mathcal F}_\mathrm{T}''' \right\rangle + \left\langle \mathcal F_\star \tilde{\mathcal F}_\mathrm{T}'' \right\rangle \left\langle \mathcal F_\mathrm{p} \tilde{\mathcal F}_\mathrm{T} \right\rangle - \left\langle \mathcal F_\star \tilde{\mathcal F}_\mathrm{T} \right\rangle \left\langle \mathcal F_\mathrm{p} \tilde{\mathcal F}_\mathrm{T}'' \right\rangle}{5\left\langle \mathcal F_\star \tilde{\mathcal F}_\mathrm{T}'' \right\rangle \left\langle \mathcal F_\star \tilde{\mathcal F}_\mathrm{T}'''\right\rangle - 2\sigma_0^2 \left(\left\langle \mathcal F_\star \tilde{\mathcal F}_\mathrm{T}'''\right\rangle^2 - \left\langle \mathcal F_\star \tilde{\mathcal F}_\mathrm{T}''\right\rangle \left\langle \mathcal F_\star \tilde{\mathcal F}_\mathrm{T}'''' \right\rangle \right)}, \quad
\hat s \simeq \frac{- \left\langle \mathcal F_\mathrm{p} \tilde{\mathcal F}_\mathrm{T}' \right\rangle + \left\langle \mathcal F_\star \tilde{\mathcal F}_\mathrm{T}''' \right\rangle \frac{\sigma^2-\sigma_0^2}{2}}{\left\langle \mathcal F_\star \tilde{\mathcal F}_\mathrm{T}''\right\rangle}.
\label{fitRV2f}
\end{equation}
Furthermore, \citet{Boue13} consider the model with only single-lined spectra,
in particular a plain Gaussian profile in $\mathcal F_\mathrm{T}$. Also, they
consider that line profiles are symmetric. This necessitates that $\left\langle \mathcal
F_\star \tilde{\mathcal F}_\mathrm{T}''' \right\rangle = 0$ and hence $\hat s$
is finally expressed by almost the same formulae in~(\ref{fitRVf})
and~(\ref{fitRV2f}), while the value of $\sigma$ becomes not important.

However, \citet{Boue13} do not mimic the procedures of
\citet{Baranne96,Pepe02} strictly. Instead of constructing the CCF
with a predefined template and subsequent fit of this CCF by a Gaussian, they assume
that the template is a fittable Gaussian itself. We follow the sequence by
\citet{Baranne96,Pepe02} more strictly, considering no fittable parameters
in $\mathcal F_\mathrm{T}$ but instead performing a Gaussian fit of the resulting
CCF. Therefore, our results for the CCF technique should not necessarily coincide with
those by \citet{Boue13} in general. Nonetheless, it is possible to
bridge them. Using eqs.~(6) from \citet{Boue13} for the case $f=0$ (out-of-transit
state) we can derive that $\langle \mathcal F_\star\mathcal F_{\mathrm
T}'' \rangle = a_0 \langle\mathcal F_\mathrm{T}
\mathcal F_\mathrm{T}''\rangle = - a_0 ||\mathcal F_\mathrm{T}'||^2 =
-a_0/(4\sigma_0^2\sqrt\pi)$. In this formulae, $a_0$ and $\sigma_0$ are the best-fit
parameters of the Gaussian template, as defined in \citep{Boue13}, and is different from
our definition. This additional relation allows us to reproduce entirely the
main formula~(12) from \citep{Boue13} work, based on our formula~(\ref{fitRVf}).

Whenever $\mathcal F_\mathrm{T}$ coincides with $\mathcal F_\star$, we
obtain from~(\ref{fitRVf}):
\begin{equation}
\hat s \simeq -\frac{\left\langle \mathcal F_{\mathrm p}\mathcal F_\star' \right\rangle}{\left\langle \mathcal F_\star\mathcal F_\star'' \right\rangle} =
  \frac{\left\langle \mathcal F_{\mathrm p}\mathcal F_\star' \right\rangle}{||\mathcal F_\star'||^2}.
\end{equation}
Taking into account all differences in the notation, this replicates eq.~(27) from
\citep{Boue13} that expresses the RM anomaly for the iodine cell technique.

Thus, our formulae allow to confidently reproduce the main results from \citep{Boue13}, but
rely on more general formulations (except for the effect of macro-turbulence that we
neglect).

\subsection{Approximations of the star and subplanet spectra}

Let us first provide the least restrictive decomposition for $\mathcal F_{\mathrm p}$:
\begin{eqnarray}
\mathcal F_{\mathrm p}(s) &=& \int\limits_{\mathcal S_{\mathrm p}} \mathcal F(s-\upsilon x) I(|\mathbfit R|,s) d\mathbfit R =
\int\limits_{\mathcal S_{\mathrm p}} \left[ \mathcal F(s-s_{\mathrm p}) + \mathcal F'(s-s_{\mathrm p}) (s_{\mathrm p}-\upsilon x) + \frac{1}{2}\mathcal F''(s-s_{\mathrm p}) (s_{\mathrm p}-\upsilon x)^2 + \ldots \right] I(|\mathbfit R|,s) d\mathbfit R = \nonumber\\
&=& \mathcal F(s-s_{\mathrm p}) M_0 + \mathcal F'(s-s_{\mathrm p}) (s_{\mathrm p} M_0-\upsilon M_1) + \frac{1}{2} \mathcal F''(s-s_{\mathrm p}) (\upsilon^2 M_2 - 2\upsilon s_{\mathrm p} M_1 + s_{\mathrm p}^2 M_0) + \ldots, \nonumber\\
& & M_k(s) = \int\limits_{\mathcal S_{\mathrm p}} x^k I(|\mathbfit R|,s)\, d\mathbfit R.
\label{Fp0}
\end{eqnarray}
Note that $M_k$ defined above are unrelated to $M_k$ defined
by \citet{Boue13}. Currently the decomposition point $s_{\mathrm p}$ is rather arbitrary, and we
still can choose it as we like. We may notice that
whenever $s_{\mathrm p}=\upsilon M_1/M_0$, the linear term in this series vanishes, leaving only the
quadratic and higher terms. Therefore, this is the natural reference point for the
decomposition. It coincides with the ``subplanet velocity'' defined in
\citep{Boue13}. Thus, we can write down
\begin{eqnarray}
\frac{\mathcal F_{\mathrm p}(s)}{M_0(s)} = \mathcal F(s-s_{\mathrm p}(s)) + \frac{1}{2} \mathcal F''(s-s_{\mathrm p}(s)) \sigma_{\mathrm p}^2(s) - \frac{1}{6} \mathcal F'''(s-s_{\mathrm p}(s)) \gamma_{\mathrm p}(s) + \mathcal O(\upsilon^4,r^4), \nonumber\\
s_{\mathrm p}(s) = \upsilon \frac{M_1(s)}{M_0(s)}, \quad
\sigma_{\mathrm p}^2(s) = \upsilon^2 \frac{M_2(s)}{M_0(s)} - s_{\mathrm p}^2(s), \quad
\gamma_{\mathrm p}(s) = \upsilon^3 \frac{M_3(s)}{M_0(s)} - 3 s_{\mathrm p}(s) \sigma_{\mathrm p}^2(s) - s_{\mathrm p}^3(s).
\label{Fp}
\end{eqnarray}
This decomposition of $\mathcal F_{\mathrm p}$ remains equally valid for the both limiting cases
introduced above, slow rotation or small planet. In both these cases, $\sigma_{\mathrm p}^2$ and
$\gamma_{\mathrm p}$ appear small. The argument $\upsilon x - s_{\mathrm p}$ in~(\ref{Fp0}) is either of the
order $\mathcal O(\upsilon)$ or $\mathcal O(r)$, where $r$ is the planet/star radii ratio.
This implies that $\sigma_{\mathrm p}^2$ is either $\mathcal O(\upsilon^2)$ or $\mathcal O(r^2)$, and
$\gamma_{\mathrm p}$ is either $\mathcal O(\upsilon^3)$ or $\mathcal O(r^3)$, and the remaining terms
of~(\ref{Fp}) are by an order higher. The quantity $s_{\mathrm p}$ is either $\mathcal
O(\upsilon)$ or $\mathcal O(r^0)$, so it is small only in the case of slow rotation.
In fact, the first two terms in~(\ref{Fp}) only reflect the effect of Doppler shift by
$s_{\mathrm p}$ and the rotational line broadening effect, characterized
by $\sigma_{\mathrm p}^2$. The third term characterizes the asymmetry effect of the rotational
broadening. Note that all characteristics $s_{\mathrm p},\sigma_{\mathrm p}^2,\gamma_{\mathrm p}$ depend
on the wavelength, due to the dependence of the limb-darkening law from the wavelength.

Till this point, we did not assume that $s_{\mathrm p}$ is small enough to justify spectra expansions
involving powers of $s_{\mathrm p}$. Now we assume that $\upsilon$ is small in comparison with the
line widths of $\mathcal F$ then hence $s_{\mathrm p}$ is small enough to perform such a
decomposition. In this case it is also legal to process the rotating star spectrum
$\mathcal F_\star$ in the way similar to~(\ref{Fp0}). Then we have
\begin{eqnarray}
\frac{\mathcal F_{\mathrm p}(s)}{M_0(s)} &=& \mathcal F(s) - \mathcal F'(s) s_{\mathrm p}(s) + \frac{1}{2} \mathcal F''(s) [\sigma_{\mathrm p}^2(s)+s_{\mathrm p}^2(s)] - \frac{1}{6} \mathcal F'''(s) [\gamma_{\mathrm p}(s)+3s_{\mathrm p}(s)\sigma_{\mathrm p}^2(s)+s_{\mathrm p}^3(s)] + \mathcal O(\upsilon^4), \nonumber\\
\frac{\mathcal F_\star(s)}{M^\star_0(s)} &=& \mathcal F(s) + \frac{1}{2} \mathcal F''(s) \sigma_\star^2(s) + \mathcal O(\upsilon^4), \qquad \sigma_\star^2(s) = \upsilon^2 \frac{M^\star_2(s)}{M^\star_0(s)}, \quad M^\star_k(s) = \int\limits_{|\mathbfit R|<1} x^k I(|\mathbfit R|,s)\, d\mathbfit R.
\label{FpFs}
\end{eqnarray}
As follows from~(\ref{FpFs}), the stellar rotation does not introduce a systematic Doppler
shift or additional asymmetry of line profiles in the uneclipsed star spectrum.

Another method to approximate these spectra is to assume that they have a simple
enough functional shape with some parameters to be defined. For example \citet{Boue13} use
extensively approximations by the Gaussian profile $\mathcal G_\beta(s-u)$. We consider
here multiline spectra, so we introduce the following multi-Gaussian function:
\begin{equation}
\mathcal G_{\bm{\beta}}(s,\mathbfit u, \mathbfit c) = \sum_{i=1}^N c_i \mathcal G_{\beta_i}(s-u_i).
\end{equation}
Whenever deemed appropriate, we may try to approximate $\mathcal F_{\mathrm p}$ or
$\mathcal F_\star$ by a function from this family.\footnote{When doing so,
we basically subtract the continuum from our spectra. As the continuum is a slowly-varying
function, in comparison with the lines, its effect on all scalar products like
$\langle\mathcal F^{(k)}\mathcal F^{(m)}\rangle$ is negligible as long as at least one
of $k$ or $m$ is nonzero. However, the continuum becomes important in the norm $||\mathcal
F||^2$, so it would be illegal to apply arbitrary normalizations
to our spectra without taking into account the continuum.} Note that in this way
of modelling all line profiles become symmetric by definition,
whereas~(\ref{Fp}) and~(\ref{FpFs}) may handle asymmetric lines too. If we approximate
$\mathcal F$ in such a way than from~(\ref{Fp}) we can obtain
\begin{eqnarray}
\frac{\mathcal F_{\mathrm p}(s)}{M_0(s)} &=& \mathcal G_{\bm{\beta}}(s-s_{\mathrm p}(s),\mathbfit u,\mathbfit c) +
\frac{1}{2} \frac{\partial^2 \mathcal G_{\bm{\beta}}}{\partial s^2}(s-s_{\mathrm p}(s),\mathbfit u,\mathbfit c) \sigma_{\mathrm p}^2(s) + \ldots =\nonumber\\
&=& \sum_{i=1}^N c_i \left[\mathcal G_{\beta_i}(s-u_i-s_{\mathrm p}(s)) + \sigma_{\mathrm p}^2(s) \frac{\partial \mathcal G_{\beta_i}}{\partial(\beta_i^2)}(s-u_i-s_{\mathrm p}(s)) \right] + \ldots =\nonumber\\
&=& \sum_{i=1}^N c_i \mathcal G_{\sqrt{\beta_i^2+\sigma_{\mathrm p}^2(s)}}(s-u_i-s_{\mathrm p}(s)) + \ldots = \mathcal G_{\bm{\beta}_{\mathrm p}(s)}(s-s_{\mathrm p}(s), \mathbfit u, \mathbfit c) + \ldots, \qquad \beta_{p,i}^2(s) = \beta_i^2+\sigma_{\mathrm p}^2(s).
\label{Fpgauss}
\end{eqnarray}
Here we applied an easy identity $\partial\mathcal G_\sigma(s)/\partial\sigma = \sigma
\partial^2\mathcal G_\sigma(s)/\partial s^2$ for each individual line profile. As we can
see, formula~(\ref{Fpgauss}) reflect nothing more than the line broadening effect by
$\sigma_{\mathrm p}^2$. The approximation~(\ref{Fpgauss}) is valid as far as the
multi-Gaussian model for $\mathcal F$ is justified, and the decomposition~(\ref{Fp})
is legal. The remaining terms in~(\ref{Fpgauss}) have the same order as in~(\ref{Fp}),
namely either $\mathcal O(\upsilon^3)$ or $\mathcal O(r^3)$. When neither $\upsilon$ nor
$r$ is small, i.e. when we deal with a large object eclipsing a fast rotating star then the
spectrum $\mathcal F_{\mathrm p}$ is not Gaussian even if $\mathcal F$ is. Moreover, its
lines might gain significant asymmetry and their shift might become different from
$s_{\mathrm p}$. Likely, this case can be only processed numerically, and we
do not consider it in our work.

Based on~(\ref{FpFs}), we may construct a similar Gaussian approximation to the star
spectrum:
\begin{equation}
\frac{\mathcal F_\star(s)}{M^\star_0(s)} = \mathcal G_{\bm{\beta}_\star}(s, \mathbfit u, \mathbfit c) + \mathcal O(\upsilon^3), \qquad
\beta_{\star,i}^2 = \beta_i^2+\sigma_\star^2,
\label{Fsgauss}
\end{equation}
but it has more restrictions than~(\ref{Fpgauss}): it is only legal for small rotation
velocities. If this is not fulfilled then $\mathcal F_\star$ is not Gaussian actually, even
if $\mathcal F$ and $\mathcal F_\mathrm{p}$ are, and to obtain $\mathcal F_\star$ we should
convolve $\mathcal F$ with a specialized rotation kernel, see e.g. \citep{Boue13}. The
result still might be approximated by a multi-Gaussian
function $\mathcal G_{\bm{\beta}_\star}(s,\mathbfit u,\mathbfit c_\star)$ with a
satisfactory accuracy:
\begin{equation}
\mathcal F_\star(s) \approx M^\star_0(s) \mathcal G_{\bm{\beta}_\star}(s, \mathbfit u, \mathbfit c_\star), \qquad
\beta_{\star,i}^2 = \beta_i^2+\sigma_\star^2
\label{Fsgauss2}
\end{equation}
but there is no guarantee that the broadening parameter $\sigma_\star$ is the here same as
defined in~(\ref{FpFs}), although it should be of the same order at least. Also, we should
introduce the best fitting values of line intensities, $\mathbfit c_\star$, which may
become somewhat different from the original $\mathbfit c$.

In our computations we often deal with various convolutions, where the following
property might be helpful:
\begin{equation}
\left\langle \mathcal G_{\beta_1}^{(k)}(s-u_1) \mathcal G_{\beta_2}^{(m)}(s-u_2) \right\rangle =
(-1)^k \mathcal G_{\sqrt{\beta_1^2+\beta_2^2}}^{(k+m)}(u_1-u_2).
\end{equation}
This identity can be proved by applying a Fourier transform to its left-hand side as
to a function of $u_1-u_2$. In particular, scalar product of two multi-Gaussian spectra can
be represented as
\begin{equation}
\left\langle \frac{\partial^k\mathcal G_{\bm{\beta}_1}}{\partial s^k}(s,\mathbfit u_1,\mathbfit c_1)
\frac{\partial^m\mathcal G_{\bm{\beta}_2}}{\partial s^m}(s,\mathbfit u_2,\mathbfit c_2) \right\rangle =
(-1)^k \sum_{i,j=1}^N c_{1i}c_{2j} \mathcal G_{\sqrt{\beta_{1i}^2+\beta_{2j}^2}}^{(k+m)}(u_{1i}-u_{2j}).
\label{prodG}
\end{equation}
In practice we often compare same or close line patterns with $\mathbfit u_1 = \mathbfit u_2$ or
$\mathbfit u_1 \approx \mathbfit u_2$. In this case~(\ref{prodG}) can be simplified further.
If all or the most of spectral lines are well separated from
each other (do not overlap), the diagonal terms of~(\ref{prodG}) are dominating, while
off-diagonal ones can be neglected:
\begin{eqnarray}
\left\langle \frac{\partial^k\mathcal G_{\bm{\beta}_1}}{\partial s^k}(s,\mathbfit u_1,\mathbfit c_1)
\frac{\partial^m\mathcal G_{\bm{\beta}_2}}{\partial s^m}(s,\mathbfit u_2,\mathbfit c_2) \right\rangle &\simeq&
(-1)^k \sum_{i=1}^N c_{1i}c_{2i} \mathcal G_{\sqrt{\beta_{1i}^2+\beta_{2i}^2}}^{(k+m)}(u_{1i}-u_{2i}) =
(-1)^k \mathcal G_{\bm{\beta}'}^{(k+m)}(0,\mathbfit u',\mathbfit c'),\nonumber\\
& & (\beta_i')^2=\beta_{1i}^2+\beta_{2i}^2, \qquad \mathbfit u'=\mathbfit u_2-\mathbfit u_1, \qquad c_i'=c_{1i}c_{2i}.
\label{prodGdiag}
\end{eqnarray}

\section{Rossiter-McLaughlin anomaly for a ``small'' rotation velocity}
\label{sec_rmobs}

We consider two types of approximations:
\begin{enumerate}
\item Small transiting planet. This means small planet/star radii ratio and small relative
subplanet flux drop $f=M_0/M_0^\star$. However, the rotation velocity $\upsilon$ and hence
the subplanet velocity $s_{\mathrm p}$ are not necessarily small and may be comparable and
even exceed the typical width of the spectral lines in $\mathcal F$. Due
to the small planet radius, the subplanet spectrum can be approximated
by~(\ref{Fp}). But the expansions~(\ref{FpFs}) are not applicable. To process this case, we
assume multiline spectra models with Gaussian line profiles,
implying representations~(\ref{Fpgauss}) and~(\ref{Fsgauss2}). We consider only first-order
approximation in $f$.

\item Small rotation velocity. This means that $\upsilon$ and $s_{\mathrm p}$ are smaller than the
typical width of the spectrum $\mathcal F$ lines. The transiting object (and hence the flux
drop $f$) is not necessarily small and can be comparable in size to the star itself. The
subplanet and rotating star spectra both can be represented via~(\ref{Fp})
and~(\ref{FpFs}).
\end{enumerate}

In this section we only give our results for the second case, because it is the case
in which our results are neat and their practical use is easy. The approximation
of the first type is considered in Appendix~\ref{sec_rmobs1}.

\subsection{Cross-correlation with a predefined template}
In the approximation of small rotation velocity we use~(\ref{FpFs}) to obtain
\begin{eqnarray}
\left\langle \mathcal F_{\mathrm p}\mathcal F_\mathrm{T}^{(k)} \right\rangle &=& \left\langle M_0 \mathcal F, \mathcal F_\mathrm{T}^{(k)}\right\rangle -
\left\langle s_{\mathrm p} M_0 \mathcal F', \mathcal F_\mathrm{T}^{(k)}\right\rangle +
\frac{1}{2} \left\langle (\sigma_{\mathrm p}^2+s_{\mathrm p}^2) M_0 \mathcal F'', \mathcal F_\mathrm{T}^{(k)}\right\rangle -
\frac{1}{6} \left\langle (\gamma_{\mathrm p}+3s_{\mathrm p}\sigma_{\mathrm p}^2+s_{\mathrm p}^3)M_0\mathcal F''', \mathcal F_\mathrm{T}^{(k)}\right\rangle +
\mathcal O(\upsilon^4), \nonumber\\
\left\langle \mathcal F_\star\mathcal F_\mathrm{T}^{(k)} \right\rangle &=& \left\langle M_0^\star \mathcal F, \mathcal F_\mathrm{T}^{(k)}\right\rangle +
\frac{1}{2}\left\langle \sigma_\star^2 M_0^\star \mathcal F'', \mathcal F_\mathrm{T}^{(k)}\right\rangle +
\mathcal O(\upsilon^4).
\end{eqnarray}
To transform these expressions to a bit more simple form, we use the property that
quantities $M_0,s_{\mathrm p},\sigma_{\mathrm p},\sigma_\star,\gamma_{\mathrm p}$ are all slowly varying functions
of wavelength, in comparison with the spectra $\mathcal F$ and $\mathcal F_\mathrm{T}$ that contains
numerous narrow lines and vary quickly. We make an additional assumption that spectral
lines are distributed more or less uniformly in the spectral range of interest and
do not reveal systematic changes of characteristics over the spectrum. In this
case variations of any selected slowly-varying function $A(s)$ are uncorrelated with
variations of $\mathcal F(s)$ and $\mathcal F_\mathrm{T}(s)$, justifying approximations of the type
\begin{equation}
\left\langle A(s), \mathcal F^{(m)}(s) \mathcal F_\mathrm{T}^{(k)}(s)\right\rangle \simeq \left(\frac{1}{s_{\max}-s_{\min}}\int\limits_{s_{\min}}^{s_{\max}} A(s) ds\right) \left\langle \mathcal F^{(m)} \mathcal F_\mathrm{T}^{(k)}\right\rangle.
\end{equation}
This implies that we can just replace the integrals $M_k(s)$ and $M_k(s)$ by their
wavelength averages $\bar M_k$ and $\bar M_k^\star$ and define averaged quantities $\bar
s_{\mathrm p},\bar\sigma_{\mathrm p},\bar\sigma_\star,\bar\gamma_{\mathrm p}$ in exactly the same manner as in~(\ref{Fp})
and~(\ref{FpFs}), but replacing $M_k$ with $\bar M_k$. For the sake of simplicity, we will
omit these averaging overlines from our further notations. Now we can write down:
\begin{eqnarray}
\left\langle \mathcal F_{\mathrm p}\mathcal F_\mathrm{T}^{(k)} \right\rangle &=& M_0 \left[ \left\langle \mathcal F, \mathcal F_\mathrm{T}^{(k)}\right\rangle -
s_{\mathrm p} \left\langle \mathcal F', \mathcal F_\mathrm{T}^{(k)}\right\rangle +
\frac{1}{2}(\sigma_{\mathrm p}^2+s_{\mathrm p}^2) \left\langle \mathcal F'', \mathcal F_\mathrm{T}^{(k)}\right\rangle -
\frac{1}{6}(\gamma_{\mathrm p}+3s_{\mathrm p}\sigma_{\mathrm p}^2+s_{\mathrm p}^3) \left\langle \mathcal F''', \mathcal F_\mathrm{T}^{(k)}\right\rangle +
\mathcal O(\upsilon^4) \right], \nonumber\\
\left\langle \mathcal F_\star\mathcal F_\mathrm{T}^{(k)} \right\rangle  &=& M_0^\star \left[ \left\langle \mathcal F, \mathcal F_\mathrm{T}^{(k)}\right\rangle +
\frac{1}{2}\sigma_\star^2 \left\langle \mathcal F'', \mathcal F_\mathrm{T}^{(k)}\right\rangle +
\mathcal O(\upsilon^4) \right].
\label{spd}
\end{eqnarray}
The wavelength averaging operation on $M_k(s)$ is equivalent to making the
same averaging on the limb-darkening law $I(R,s)$. The limb-darkening is usually
represented as a linear combination of several simple functional terms that are independent
of the wavelength, but their coefficients are. Therefore, such an averaging can be reduced
to an averaging of only the limb-darkening coefficients in $I(R,s)$.

Now, using formulae~(\ref{fitRV}) and~(\ref{spd}), and the constraint~(\ref{ccfstar}),
we can finally derive an approximation of the RM anomaly:
\begin{eqnarray}
\hat s &=& V_1  + \nu V_2 + \mu V_3 + \mathcal O(\upsilon^4), \qquad
\nu = \frac{1}{2}\frac{\langle\mathcal F \mathcal F_\mathrm{T}'''\rangle}{\langle\mathcal F \mathcal F_\mathrm{T}''\rangle}, \quad
\mu = -\frac{1}{6}\frac{\langle\mathcal F \mathcal F_\mathrm{T}''''\rangle}{\langle\mathcal F \mathcal F_\mathrm{T}''\rangle}, \nonumber\\
& & V_1 = -\frac{f s_{\mathrm p}}{1-f} = - \frac{M_1\upsilon}{M_0^\star-M_0}, \nonumber\\
& & V_2 =  \frac{f}{1-f} \left(\sigma_\star^2-\sigma_{\mathrm p}^2-\frac{s_{\mathrm p}^2}{1-f}\right) = \frac{\upsilon^2}{M_0^\star-M_0} \left( \frac{M_0}{M_0^\star} M_2^\star - M_2 - \frac{M_1^2}{M_0^\star - M_0} \right) , \nonumber\\
& & V_3 = \frac{f}{1-f} \left(\gamma_{\mathrm p} + 3 s_{\mathrm p} \frac{\sigma_{\mathrm p}^2-\sigma_\star^2}{1-f} + s_{\mathrm p}^3 \frac{1+f}{(1-f)^2}\right) = \frac{\upsilon^3}{M_0^\star-M_0}
\left( M_3 - 3 M_1 \frac{M_2^\star-M_2}{M_0^\star-M_0} + \frac{2 M_1^3}{(M_0^\star-M_0)^2} \right).
\label{RMtmpl}
\end{eqnarray}
The coefficients $\nu$ and $\mu$ only depend on the star spectrum and on the
cross-correlation template. They do not depend on the transit geometry and do not vary
during the transit. The quantities that vary during the transit are $V_k$.

It follows from~(\ref{ccf2}) that in the case when the CCF is fitted by a Gaussian, we may
use the same formulae for $\hat s$ as in the case of a
direct CCF minimization, but replacing $\mathcal F_\mathrm{T}(s)$ by its gaussian-broadened
convolution $\tilde{\mathcal F}_\mathrm{T}(s|\sigma)$. The broadening parameter $\sigma$
should be set to the best fitting value, defined by~(\ref{sigma}) or
approximated in~(\ref{fitRV2}). Here we must take care of the mutual correlation dependence
between $\sigma$ and $\hat s$. In the formulae~(\ref{RMtmpl}), the
template $\mathcal F_\mathrm{T}$ is present in the coefficients $\nu$ and $\mu$,
so via $\sigma$ they become also dependent on the transit geometry and phase as $V_k$.
To solve the task rigorously, we should decompose both $\sigma$ and $\hat s$ into powers of
$\upsilon$, solving two equations jointly. Fortunately, such a complicated procedure
becomes unnecessary. Substituting~(\ref{spd}) in~(\ref{fitRV2}) and~(\ref{ccfstar2}) it can
be easily obtained that the first-order $\mathcal O(\upsilon)$ term in $\sigma$ actually
vanishes, so that $\sigma = \sigma_0 + \mathcal O(\upsilon^2)$. This means that in the
definitions of $\nu$ and $\mu$ we may just replace $\mathcal F_\mathrm{T}(s)$ by
$\tilde{\mathcal F}_\mathrm{T}(s|\sigma_0)$, which does not depend on the transit geometry
again. This would introduce an additional error in~(\ref{RMtmpl}) of $\mathcal
O(\upsilon^4)$, which we neglect anyway. Therefore, formulae~(\ref{RMtmpl}) remain
almost the same for the both flavours of the cross-correlation technique.

To perform a fitting of RV data with the model~(\ref{RMtmpl}), we likely need to compute
partial derivatives with respect its parameters, which are necessary for
gradient minimization of the chi-square function or other goodness-of-fit statistic.
Therefore, we simultaneously give expressions for partial derivatives of $V_k$
over $M_k$ and $M_k^\star$:
\begin{eqnarray}
V_1 &:&\quad -\frac{\partial V_1}{\partial M_0} = \frac{\partial V_1}{\partial M_0^\star} = \frac{M_1\upsilon}{(M_0^\star-M_0)^2} = \frac{1}{M_0^\star}\frac{f s_\mathrm{p}}{(1-f)^2}, \qquad
\frac{1}{\upsilon} \frac{\partial V_1}{\partial M_1} = - \frac{1}{M_0^\star-M_0} = -\frac{1}{M_0^\star}\frac{1}{1-f}, \nonumber\\
V_2 &:&\quad \frac{\partial V_2}{\partial M_0} = \frac{\upsilon^2}{(M_0^\star-M_0)^2}\left(M_2^\star-M_2-\frac{2M_1^2}{M_0^\star-M_0}\right) =
 \frac{1}{M_0^\star} \frac{1}{(1-f)^2} \left( \sigma_\star^2 - f\sigma_\mathrm{p}^2 - f s_\mathrm{p}^2\frac{1+f}{1-f} \right), \nonumber\\
 & &\quad \frac{\partial V_2}{\partial M_0^\star} = -\frac{\partial V_2}{\partial M_0} + \upsilon^2\frac{M_2^\star}{{M_0^\star}^2} = -\frac{\partial V_2}{\partial M_0} + \frac{\sigma_\star^2}{M_0^\star}, \qquad
 \frac{\partial V_2}{\partial M_1} = 2\upsilon \frac{\partial V_1}{\partial M_0}, \qquad
 \frac{\partial V_2}{\partial M_2} = \upsilon \frac{\partial V_1}{\partial M_1}, \nonumber\\
 & &\quad \frac{1}{\upsilon^2}\frac{\partial V_2}{\partial M_2^\star} = \frac{M_0}{(M_0^\star-M_0)M_0^\star} = \frac{1}{M_0^\star}\frac{f}{1-f}, \nonumber\\
V_3 &:&\quad \frac{\partial V_3}{\partial M_0} = -\frac{\partial V_3}{\partial M_0^\star} = \frac{\upsilon^3}{(M_0^\star-M_0)^2}\left(M_3 - 6 M_1 \frac{M_2^\star-M_2}{M_0^\star-M_0} - \frac{6M_1^3}{(M_0^\star-M_0)^2} \right) =\nonumber\\
& &\quad \phantom{\frac{\partial V_3}{\partial M_0} = -\frac{\partial V_3}{\partial M_0^\star}} =\frac{1}{M_0^\star}\frac{f}{(1-f)^2}\left(\gamma_\mathrm{p} + 3 s_\mathrm{p} \frac{(1+f)\sigma_\mathrm{p}^2-2\sigma_\star^2}{1-f} + s_\mathrm{p}^3 \frac{1+4f+f^2}{(1-f)^2}\right), \nonumber\\
& &\quad \frac{\partial V_3}{\partial M_1}=-3\upsilon\frac{\partial V_2}{\partial M_0}, \qquad \frac{\partial V_3}{\partial M_2} = 3\upsilon^2 \frac{\partial V_1}{\partial M_0^\star}, \qquad \frac{\partial V_3}{\partial M_2^\star} = 3\upsilon^2 \frac{\partial V_1}{\partial M_0}, \qquad \frac{\partial V_3}{\partial M_3} = -\upsilon^2 \frac{\partial V_1}{\partial M_1}.
\label{Vkdiff}
\end{eqnarray}
The momenta $M_k$ with their partial derivatives will be computed in the following sections
of the paper.

Although the coefficients $\nu$ and $\mu$ in~(\ref{RMtmpl}) are expressed by
explicit formulae here, we believe that in practice it is difficult to predict
them reliably, especially in those works where a reanalysis of public releases
of Doppler data is performed, and authors do not have access to the full
internal characteristics of the Doppler reduction pipeline. In such a case, we suggest
to treat $\nu$ and $\mu$ as additional free parameters of the RV curve fit, similarly to
e.g. the limb-darkening coefficients. In this case, three fittable coefficients of the
decomposition should better be defined as
\begin{equation}
\upsilon' = \upsilon c = V\sin i, \qquad \nu' = \nu \upsilon^2 c = \nu \upsilon V\sin i, \qquad \mu' = \mu \upsilon^3 c = \mu \upsilon^2 V \sin i,
\end{equation}
because $V_k$ imbed the power factors $\upsilon^k$, and in practice we measure unnormalized
Doppler shift $v_z$ rather than $v_z/c$. The
quantities $\nu\upsilon$ and $\mu\upsilon^2$ are adimensional, while $\nu'$ and $\mu'$ have
the dimension of velocity.

Note that $\nu$ and $\mu$ in~(\ref{RMtmpl}) are defined via the surface spectrum $\mathcal
F$, which is not observable in practice. Here it is admissible to substitute the observable
$\mathcal F_\star$ spectrum in place of $\mathcal F$, because the error caused in $\hat s$
by such a substitution is only $\mathcal O(\upsilon^4)$, which is neglected anyway.

Recall that the template $\mathcal F_\mathrm{T}$ should be shifted in such a way as
to satisfy~(\ref{ccfstar}), which in the approximation of small $\upsilon$ turnes to
\begin{equation}
\langle\mathcal F \mathcal F_\mathrm{T}'\rangle+\frac{\sigma_\star^2}{2} \langle\mathcal F \mathcal F_\mathrm{T}'''\rangle + \mathcal O(\upsilon^4) = 0.
\label{ccfsa}
\end{equation}

We expect that for any reasonably chosen template the value of $\mu$ should be positive.
This is because we can also rewrite $6\mu = \langle \mathcal F'' \mathcal F_\mathrm{T}''
\rangle/\langle \mathcal F' \mathcal F_\mathrm{T}' \rangle$, and if $\mathcal F_\mathrm{T}$
and $\mathcal F$ have lines in the same or close positions then the both products of the
derivatives remain positive over the most of the wavelengths range. The value of $\mu$ can
be negative only if the lines positions in $\mathcal F_\mathrm{T}$ have little common with
those in $\mathcal F$, or e.g. if there are many emission lines that are wrongly
modelled as absorption ones.

In other words, if the value of $\mu$ determined from the observations of the RM
effect appeared negative in a particular case, this indicates that something is wrong with
our model, rendering it unreliable. The coefficient $\nu$ may have any sign, however.

Non-zero $\nu$ may indicate either an imperfect match of the cross-correlation
template lines with those in the star spectrum, or systematic asymmetry of the
line profiles. The latter fact is of a high importance, because this means that asymmetric
line profiles may require additional correction of the RM curve exceeding, and the order of
this correction is larger than of the corrections considered by \citet{Hirano10} and
\citet{Boue13}. The quanitity $1/\sqrt\mu$ is a characteristic of an averaged width of line
profiles. For example, for multi-Gaussian $\mathcal F$ and $\mathcal F_\mathrm{T}$
(see App.~\ref{sec_rmobs1} for definitions) we have
\begin{eqnarray}
\nu &\simeq& \frac{1}{2}\frac{\sum_{i=1}^N c_i c_{\mathrm{T},i} \mathcal G_{\sqrt{\beta_i^2+\beta_{\mathrm{T},i}^2}}'''(-\Delta u_i)}{\sum_{i=1}^N c_i c_{\mathrm{T},i} \mathcal G_{\sqrt{\beta_i^2+\beta_{\mathrm{T},i}^2}}''(-\Delta u_i)} =
\frac{3}{2} \frac{\sum_{i=1}^N \frac{c_i c_{\mathrm{T},i}}{(\beta_i^2+\beta_{\mathrm{T},i})^{5/2}} \Delta u_i}{\sum_{i=1}^N \frac{c_i c_{\mathrm{T},i}}{(\beta_i^2+\beta_{\mathrm{T},i}^2)^{3/2}}} + \mathcal O(\Delta\mathbfit u^2), \nonumber\\
\mu &\simeq& - \frac{1}{6}\frac{\sum_{i=1}^N c_i c_{\mathrm{T},i} \mathcal G_{\sqrt{\beta_i^2+\beta_{\mathrm{T},i}^2}}''''(-\Delta u_i)}{\sum_{i=1}^N c_i c_{\mathrm{T},i} \mathcal G_{\sqrt{\beta_i^2+\beta_{\mathrm{T},i}^2}}''(-\Delta u_i)} =
\frac{1}{2} \frac{\sum_{i=1}^N \frac{c_i c_{\mathrm{T},i}}{(\beta_i^2+\beta_{\mathrm{T},i})^{5/2}}}{\sum_{i=1}^N \frac{c_i c_{\mathrm{T},i}}{(\beta_i^2+\beta_{\mathrm{T},i}^2)^{3/2}}} +\mathcal O(\Delta\mathbfit u^2),
\label{numuG}
\end{eqnarray}
while the constraint~(\ref{ccfsa}) can be reduced to
\begin{equation}
\sum_{i=1}^N \frac{c_i c_{\mathrm{T},i}}{(\beta_i^2+\beta_{\mathrm{T},i}^2)^{3/2}} \Delta u_i +\mathcal O(\upsilon^2)+\mathcal O(\Delta\mathbfit u^2) = 0.
\end{equation}
Thus, in the approximation of Gaussian profiles, the quanitiy $1/(2\mu)$ characterizes
an average value for $\beta_i^2+\beta_{\mathrm{T},i}^2$.

\subsection{Cross-correlation with an out-of-transit stellar spectrum or
parametric modelling of the stellar spectrum (iodine cell technique)}

Now we should just substitute $\mathcal F_\star$ in place of $\mathcal F_\mathrm{T}$ in the
formulae presented above. Formulae~(\ref{RMtmpl}) can be transformed to the following:
\begin{equation}
\hat s = V_1 + \mu V_3 + \mathcal O(\upsilon^4), \qquad \mu = \frac{1}{6} \frac{||\mathcal F''||^2}{||\mathcal F'||^2}.
\label{RMidn}
\end{equation}
We can see that now the term with $V_2$, which was responsible for either template
imperfections or asymmetry of spectral lines, disappeared. The coefficient $\mu$
is now guaranteedly positive. As before, we may treat $\mu$ as a fittable parameter of the
model, if we do not have enough knowledge of the spectra details. We however have a concern
that in practice a subtle violation of our simplificating assumptions may cause additional
disturbing effects in $\hat s$. Therefore, it might appear reasonable to use in practice
the full three-term formula~(\ref{RMtmpl}) even for Doppler data obtained with the
iodine cell technique. At least, it might be a matter of practical verification with
real data, whether the term with $V_2$ indeed becomes negligible in this case.

For multi-Gaussian $\mathcal F$ we obtain
\begin{equation}
\mu \simeq \frac{1}{4} \left.\left(\sum_{i=1}^N \frac{c_i^2}{\beta_i^5}\right)\right/\left(\sum_{i=1}^N \frac{c_i^2}{\beta_i^3}\right).
\label{muG}
\end{equation}
In this approximation, $1/(4\mu)$ measures an average value for $\beta_i^2$.

\section{Computing the RV momenta}
\label{sec_mom}

\begin{figure*}
\includegraphics[width=0.80\linewidth]{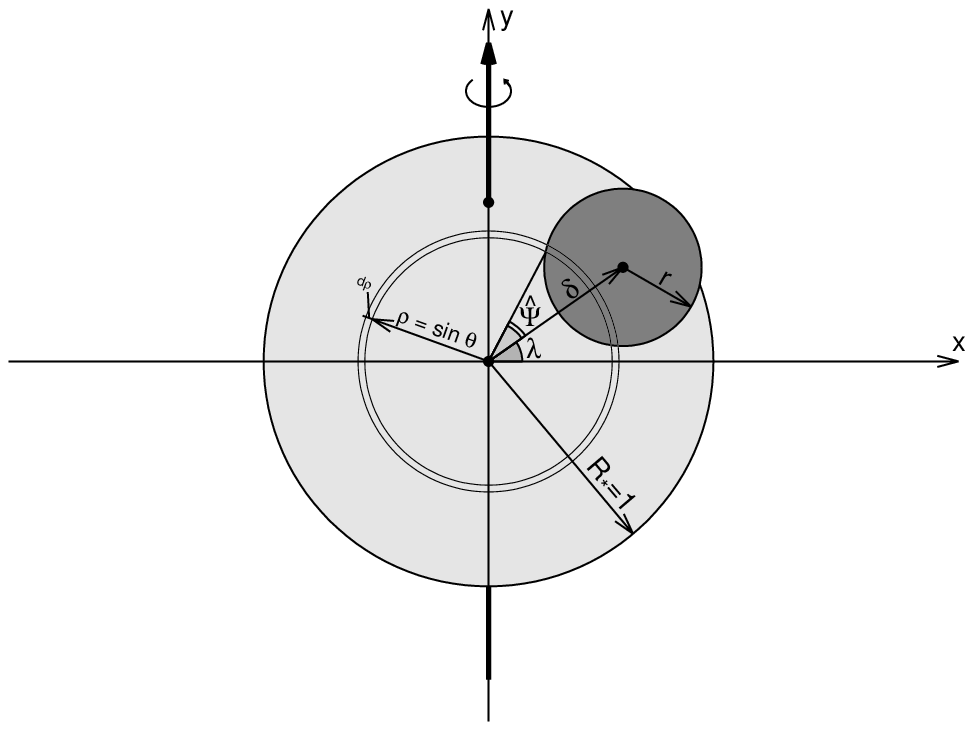}
\caption{Geometry of the transit illustrating the Rossiter-McLaughling effect.}
\label{fig_tsch}
\end{figure*}

We need to compute the integral momenta
\begin{equation}
M_k(\delta,r,\lambda) = \int\limits_{S(\delta,r,\lambda)} x^k I(|\mathbfit R|)\, d\mathbfit R, \qquad
M_k^\star = \left. M_k\right|_{\delta=0,r=1} = \int\limits_{|\mathbfit R|<1} x^k I(|\mathbfit R|)\, d\mathbfit R \qquad k=0,1,2,3.
\end{equation}
Here we consider them as functions of three parameters $\delta,r,\lambda$. Their definition
and geometrical layout are given in Fig.~\ref{fig_tsch}. The star radius is assumed unit.

To compute these momenta, we apply an adaptation of the approach developed
by \citet{AbubGost13} for transit lightcurve modeling. Define auxiliary functions
\begin{equation}
\mathcal A(x) = \Re\arccos x = \left\{\begin{array}{@{}l@{\;}l@{}}\pi, & x<-1, \\ \arccos x, & |x|<1, \\ 0, & x>1, \end{array}\right. \quad
\mathcal Q(x) = \Re\sqrt x = \left\{\begin{array}{@{}l@{\;}l@{}}\sqrt x, & x>0, \\ 0, & x<0, \end{array}\right. \quad
\frac{d\mathcal A}{dx} = - \mathcal Q\left(\frac{1}{1-x^2}\right), \quad
\Psi(\delta,x,y) = \mathcal A \left(\frac{\delta^2+x^2-y^2}{2x\delta}\right).
\end{equation}
Then, extending formula~(7) from \citep{AbubGost13} to take into account additional factor
$x^k$, we can write down:
\begin{equation}
M_k = \int\limits_0^1 I(\rho) \rho^{k+1} d\rho \int\limits_{-\Psi(\delta,\rho,r)}^{\Psi(\delta,\rho,r)} \cos^k (\varphi+\lambda)\, d\varphi.
\end{equation}
\citet{AbubGost13} computed $M_0$ and its derivatives with respect
to $r$ and $\delta$, which are necessary to express the flux reduction $\Delta
L$ and its parametric gradient (for further fitting of the model by a gradient descent).
Here we also compute $M_k$ with their derivatives for $k=1,2,3$.

Let us put $\rho=\sin\theta$. The limb-darkening law is often modelled by a polynomial
in $\cos\theta$. Then $M_k$ can be expressed via linear combinations
of the following integrals
\begin{equation}
H_{nk} = \int\limits_0^{\frac{\pi}{2}} \sin^{k+1}\theta \cos^{n+1}\theta\, d\theta \int\limits_{\lambda-\hat\Psi}^{\lambda+\hat\Psi} \cos^k\varphi\, d\varphi, \qquad \hat\Psi=\Psi(\delta,\sin\theta,r), \nonumber\\
\end{equation}
Assuming at first that $k$ is odd, rewrite $\cos^k\varphi$ as a trigonometric sum of
multiple argument and compute the inner integral:
\begin{eqnarray}
H_{nk} &=& \int\limits_0^{\frac{\pi}{2}} \sin^{k+1}\theta \cos^{n+1}\theta\, d\theta \int\limits_{\lambda-\hat\Psi}^{\lambda+\hat\Psi} \frac{1}{2^{k-1}}\sum_{j=0}^{[k/2]} C_k^j \cos(k-2j)\varphi\, d\varphi = \nonumber\\
       &=& \frac{1}{2^{k-2}} \int\limits_0^{\frac{\pi}{2}} \sin^{k+1}\theta \cos^{n+1}\theta \sum_{j=0}^{[k/2]} \frac{C_k^j}{k-2j} \sin(k-2j)\hat\Psi\, \cos(k-2j)\lambda \, d\theta = \nonumber\\
       &=& \frac{1}{2^{k-2}} \sum_{j=0}^{[k/2]} \frac{C_k^j}{k-2j} \cos(k-2j)\lambda \int\limits_0^{\frac{\pi}{2}} \sin^{k+1}\theta\,\cos^{n+1}\theta\,U_{k-2j-1}(\cos\hat\Psi) \sin\hat\Psi \,d\theta.
\end{eqnarray}
where $U_n$ are Chebyshev polynomials of the second kind. If $k$ is even, then the
procedure is similar, but the sum should also contain a term with $j=k/2$, which must be
handled separately due to a degeneracy. In general, we can represent
$H_{nk}$ via the following trigonometric polynomial in $\lambda$:
\begin{equation}
H_{nk} = \sum_{j=0}^{[k/2]} \mathcal I_{nkj} \cos(k-2j)\lambda,
\label{Hnk}
\end{equation}
where
\begin{equation}
\mathcal I_{nkj} = \frac{1}{2^{k-2}} \frac{C_k^j}{k-2j} \int\limits_0^{\frac{\pi}{2}} \sin^{k+1}\theta\,\cos^{n+1}\theta\,U_{k-2j-1}(\cos\hat\Psi) \sin\hat\Psi \,d\theta, \quad 2j<k, \qquad
\mathcal I_{n,2j,j} = \frac{C_{2j}^j}{2^{2j-1}} \int\limits_0^{\frac{\pi}{2}} \sin^{2j+1}\theta \cos^{n+1}\theta\, \hat\Psi\, d\theta.
\end{equation}
By extracting from $\cos^{n+1}\theta$ a multiplier $\cos^2\theta=1-\sin^2\theta$, we can
obtain the following recurrent relation:
\begin{equation}
\mathcal I_{nkj} = \mathcal I_{n-2,k,j} - 4\frac{(j+1)(k-j+1)}{(k+1)(k+2)} \mathcal I_{n-2,k+2,j+1}, \qquad 2j\leq k.
\label{Inkjrec}
\end{equation}
This can be used to reduce the index $n$ to either $n=0$ or $n=1$, by the cost
of increasing $k$ and $j$. In practice we only need this formula to reduce $n=2$ (quadratic
limb-darkening term) to $n=0$.

Now let us define
\begin{equation}
a=1-(\delta-r)^2, \quad a(\theta)=a-\cos^2\theta, \qquad b=(\delta+r)^2-1, \quad b(\theta) = b+\cos^2\theta.
\end{equation}
This yields
\begin{equation}
\cos\hat\Psi \sin\theta = (\delta-r) + \frac{a(\theta)}{2\delta}, \qquad \sin\hat\Psi \sin\theta = \frac{\mathcal Q[a(\theta)b(\theta)]}{2\delta}.
\end{equation}
In fact, the integration range can be limited to only the range where $\sin\hat\Psi\neq 0$,
corresponding to $-b\leq \cos^2\theta \leq a\leq 1$. For convenience, we need to transform
this variable range to a constant one. The case $a<0$ is trivial, as $\sin\hat\Psi\equiv 0$
everywhere. In the case $a>0$ we can introduce the following replacement
$\theta\mapsto t$:
\begin{equation}
\cos^2\theta = a - \min(a,a+b) t^2, \quad
\sin\theta\cos\theta\, d\theta = \min(a,a+b)\, t\, dt.
\label{subs}
\end{equation}
After this replacement, the new integration variable $t$ always spans the same segment
$[0,1]$. By making this replacement, expanding Chebyshev polynomials
as $U_n(x) = \sum_{l=0}^{[n/2]} u_{nl} x^{n-2l}$, and introducing additional auxiliary
designations, we may rewrite the integrals $\mathcal I_{nkj}$ as follows:
\begin{eqnarray}
\mathcal I_{nkj} &=& \frac{(1-q^2)^{\frac{n}{2}}}{2^{k-2}} \frac{C_k^j}{k-2j} \frac{p}{2\delta} \int\limits_0^1
 P_{k-j}\left(pt^2\right) \sqrt{\max(1,m)-t^2}\, \left(1-\min(m,1)t^2\right)^{\frac{n}{2}}\, dt, \qquad 2j<k, \nonumber\\
P_{k-j}(x) &=& x \left(q^2+x\right)^{\frac{k-1}{2}} U_{k-2j-1}\left(\frac{q + \frac{x}{2\delta}}{\sqrt{q^2+x}}\right) = x \sum_{l=0}^{\left[\frac{k-1}{2}\right]-j} u_{k-2j-1,l}
 \left(q + \frac{x}{2\delta}\right)^{k-2j-2l-1} \left(q^2+x\right)^{j+l} \nonumber\\
& & q=\delta-r, \qquad m=\frac{4r\delta}{1-q^2}, \qquad p=\min(4r\delta,1-q^2)=4r\delta\min\left(1,\frac{1}{m}\right)=(1-q^2)\min(1,m).
\label{Inkj}
\end{eqnarray}
Note that although there is a division by $\delta$ in $p/(2\delta)$ and $x/(2\delta)$,
in actuality there is no pecularity at $\delta=0$, because by definition
$p<4r\delta$ and thus $p/(2\delta)<2r$. Note that $0<m<1$ corresponds to full phase of a
transit ($r+\delta<1$), while $m>1$ corresponds to a partial occultation
$|r-\delta|<1<r+\delta$. The cases of a full eclipse or no eclipse
($|r-\delta|>1$) would correspond to $m<0$, which are illegal in these formulae by
definition. In the latter case, $\mathcal I_{nkj}=0$ for $2j<k$.

The case $k=2j$ is more complicated due to a ``naked'' $\hat\Psi$ in the integrad.
We may apply two ways of integration by parts:
\begin{eqnarray}
\mathcal I_{n,2j,j} &=& -\frac{C_{2j}^j}{2^{2j-1}} \int\limits_0^{\frac{\pi}{2}} \hat\Psi \sin^{2j}\theta\, \frac{d\cos^{n+2}\theta}{n+2} =
\delta_{0j}\frac{2\pi\Theta(-q)}{n+2} + \frac{2j-1}{n+2} \mathcal I_{n+2,2j-2,j-1} + \frac{1}{n+2}\frac{C_{2j}^j}{2^{2j-1}}\int\limits_0^{\frac{\pi}{2}} \sin^{2j}\theta \cos^{n+2}\theta \frac{d\hat\Psi}{d\theta} d\theta = \nonumber\\
&=& \frac{C_{2j}^j}{2^{2j-1}} \int\limits_0^{\frac{\pi}{2}} \hat\Psi \cos^n\theta\, \frac{d\sin^{2j+2}\theta}{2j+2} =
\delta_{0n} \frac{C_{2j}^j}{2^{2j}} \frac{\Psi(\delta,1,r)}{j+1} + \frac{n}{2j+1} \mathcal I_{n-2,2j+2,j+1} - \frac{1}{j+1} \frac{C_{2j}^j}{2^{2j}}\int\limits_0^{\frac{\pi}{2}} \sin^{2j+2}\theta \cos^n\theta \frac{d\hat\Psi}{d\theta} d\theta,
\label{rtmp}
\end{eqnarray}
where $\delta_{ik}$ is Kronecker delta (not to be mixed with the planet-star
distance $\delta$, which is unindexed), and we use that $\lim_{\rho\to +0}
\Psi(\delta,\rho,r) = \pi \Theta(-q)$ (with $\Theta$ being the Heaviside
function). Taking into account~(\ref{Inkjrec}), both formulae~(\ref{rtmp}) appear to yield
the same recurrent relation, if $n\neq 0$:
\begin{equation}
(n+2j+2) \mathcal I_{n,2j,j} = (2j-1) \mathcal I_{n,2j-2,j-1} + G_{n+2,j}, \quad j>0, \qquad
G_{nj} = \frac{C_{2j}^j}{2^{2j-1}}\int\limits_0^{\frac{\pi}{2}} \sin^{2j}\theta \cos^n \theta \frac{d\hat\Psi}{d\theta} d\theta.
\label{Idrec}
\end{equation}
For $n=0$ this relation is still valid thanks to the first of eq.~(\ref{rtmp}), but we also
obtain from the second eq. of~(\ref{rtmp}) an independent non-recurrent formula
\begin{equation}
\mathcal I_{0,2j,j} = \frac{C_{2j}^j}{2^{2j}} \frac{\Psi(\delta,1,r)}{j+1} - \frac{G_{0,j+1}}{2j+1}, \quad j\geq 0.
\label{Ide}
\end{equation}
It can be verified by direct substitution that this formula satisfies~(\ref{Idrec}). The
recursion~(\ref{Idrec}) can be used to reduce $j$ until we meet $j=0$. For $j=0$ we may use
one of the following schemes:
\begin{equation}
(n+2)\mathcal I_{n,0,0} = 2\pi \Theta(-q) + G_{n+2,0}, \qquad
\mathcal I_{0,0,0} = \Psi(\delta,1,r) - G_{0,1}.
\label{Id0}
\end{equation}
We use the second formula of~(\ref{Id0}) for $n=0$, because this simplifies some
computations.

To compute $G_{nj}$, we express the derivative of $\hat\Psi$
by differentiating $\cos\hat\Psi$ and then apply the substitution~(\ref{subs}):
\begin{equation}
G_{nj} = r (1-q^2)^{\frac{n}{2}} \frac{C_{2j}^j}{2^{2j-2}} \int\limits_0^1 \left(q - \frac{p}{2r} t^2\right) \left(q^2+pt^2\right)^{j-1} \frac{\left(1-\min(1,m)t^2 \right)^{\frac{n}{2}}}{\sqrt{\max(1,m)-t^2}}\, dt.
\label{Id}
\end{equation}
Note that $p/(2r)<2\delta$, so there is no pecularity at $r=0$. Note that in the case of a
full eclipse or no eclipse, $|r-\delta|>1$, we have $m<0$ and put $G_{nj}=0$ by definition.

Now let us review the results. Our final useable
formulae are~(\ref{Hnk},\ref{Inkjrec},\ref{Inkj},\ref{Idrec},\ref{Id0},\ref{Id}). For even
$n$ all integrals are elementary, because their integrads are rational functions of $t$ and
of the radical $\sqrt{\max(m,1)-t^2}$. For odd $n$, the integrands are
rational functions of $t$ and $\sqrt{(1-t^2)(1-\min(m,1/m)t^2)}$, implying
that all integrals are elliptic and can be expressed via Legendre complete
elliptic integrals with the parameter $\min(m,1/m)$. In the most cases, this should be only
the Legendre integrals of the first and second kind, and only in $G_{1,0}$ we meet the
Legendre integral of the third kind, where $t$ is present in a denominator
of the integrand's rational part. This integral, however, affects $\mathcal I_{1,2j,j}$ for
all $j$, due to the recursion~(\ref{Idrec}).

Our approach allows to give exact and explicit formulae for all velocity momenta for any
polynomial limb-darkening law. However, in this work we only need momenta up to cubic order
($k=0,1,2,3$) and up to quadratic limb-darkening law ($n=0,1,2$). From now on, we stop
using generic notations and focus on the computation of $H_{nk}$ for the specified indices
$n$ and $k$.

Consider the quadratic limb-darkening model
\begin{eqnarray}
I(\rho) &=& I^\mathrm{c}(\rho) - \Lambda_l I^\mathrm{l}(\rho) - \Lambda_q I^\mathrm{q}(\rho) = (1-\Lambda_l-2\Lambda_q) I^\mathrm{c} + (\Lambda_l+2\Lambda_q) (I^\mathrm{c}-I^\mathrm{l}) + \Lambda_q (2I^\mathrm{l}-I^\mathrm{q}), \nonumber\\
& & I^\mathrm{c}=1, \qquad I^\mathrm{l} = 1-\cos\theta, \quad I^\mathrm{c}-I^\mathrm{l} = \cos\theta, \qquad I^\mathrm{q} = (1-\cos\theta)^2, \quad 2I^\mathrm{l}-I^\mathrm{q} = 1-\cos^2\theta.
\label{Ild}
\end{eqnarray}
Therefore, using~(\ref{Inkjrec}) for terms with $n=2$, we have:
\begin{equation}
\begin{array}{lll}
\multicolumn{3}{l}{M_k = (1-\Lambda_l-2\Lambda_q) M_k^\mathrm{c} + (\Lambda_l+2\Lambda_q) (M_k^\mathrm{c}-M_k^\mathrm{l}) + \Lambda_q (2M_k^\mathrm{l}-M_k^\mathrm{q}), \qquad k=0,1,2,3,} \\
M_0^\mathrm{c} = \mathcal I_{0,0,0}, & M_0^\mathrm{c}-M_0^\mathrm{l} = \mathcal I_{1,0,0}, & 2M_0^\mathrm{l}-M_0^\mathrm{q} = 2 \mathcal I_{0,2,1}, \\
M_1^\mathrm{c} = \mathcal I_{0,1,0} \cos\lambda, & M_1^\mathrm{c}-M_1^\mathrm{l} = \mathcal I_{1,1,0} \cos\lambda, & 2M_1^\mathrm{l}-M_1^\mathrm{q} = \frac{4}{3} \mathcal I_{0,3,1} \cos\lambda, \\
M_2^\mathrm{c} = \mathcal I_{0,2,0} \cos 2\lambda + \mathcal I_{0,2,1}, & M_2^\mathrm{c}-M_2^\mathrm{l} = \mathcal I_{1,2,0} \cos 2\lambda + \mathcal I_{1,2,1}, & 2M_2^\mathrm{l}-M_2^\mathrm{q} = \mathcal I_{0,4,1} \cos 2\lambda + \frac{4}{3} \mathcal I_{0,4,2}, \\
M_3^\mathrm{c} = \mathcal I_{0,3,0} \cos 3\lambda + \mathcal I_{0,3,1} \cos\lambda, & M_3^\mathrm{c}-M_3^\mathrm{l} = \mathcal I_{1,3,0} \cos 3\lambda + \mathcal I_{1,3,1} \cos\lambda, & 2M_2^\mathrm{l}-M_2^\mathrm{q} = \frac{4}{5} \mathcal I_{0,5,1} \cos 3\lambda + \frac{6}{5} \mathcal I_{0,5,2} \cos\lambda.
\end{array}
\label{Mk}
\end{equation}
Thus, we need to compute $16$ integrals $\mathcal I_{nkj}$ in total: $11$ of them are of
type $k>2j$ (with $7$ elementary, and $4$ elliptic) and $5$ are of type $k=2j$
($3$ elementary, $2$ elliptic). Note that
\begin{equation}
\mathcal I_{0,0,0} = \Psi(\delta,1,r)-G_{0,1}, \quad
\mathcal I_{0,2,1} = \frac{\mathcal I_{0,0,0}+G_{2,1}}{4}, \quad
\mathcal I_{0,4,2} = \frac{3\mathcal I_{0,2,1}+G_{2,2}}{6}, \quad
\mathcal I_{1,0,0} = \frac{2\pi\Theta(-q) + G_{3,0}}{3}, \quad
\mathcal I_{1,2,1} = \frac{\mathcal I_{1,0,0} + G_{3,1}}{5}.
\end{equation}

The remaining part is to compute integrals~(\ref{Inkj},\ref{Id}) for the indices specified
above. This is a routine but difficult work due to quickly growing formulae. We used MAPLE
computer algebra to compute these integrals in a symbolic form. The results are given in
Table~\ref{tab_Ik}. We represent all $\mathcal I_{nkj}$ and their derivatives as
linear combinations of the following functions: elementary $\psi,\varphi$, and $Q$ for
$n=0,2$ and elliptic $\tilde K,\tilde E$, and $\tilde\Pi$ (or $\tilde J$) for $n=1$. The
coefficients are algebraic polynomials in $r$ and $\delta$. The degree of these polynomials
may reach $8$, and we tried to reduce them by reusing an auxiliary function $W$,
whenever possible. All these analytic expressions were verified by comparison with numeric
calculation of the integrals.

In this work we adopt the following definition of the complete elliptic integrals (Legendre
normal forms):
\begin{equation}
K(m) = \int\limits_0^{\frac{\pi}{2}} \frac{d\theta}{\sqrt{1-m\sin^2\theta}}, \qquad
E(m) = \int\limits_0^{\frac{\pi}{2}} \sqrt{1-m\sin^2\theta}\,d\theta, \qquad
\Pi(n|m) = \int\limits_0^{\frac{\pi}{2}} \frac{d\theta}{(1+n\sin^2\theta) \sqrt{1-m\sin^2\theta}}.
\end{equation}
Note that the sign of $n$ here is opposite to what is adopted by \citep{AbubGost13} and by
MAPLE, but it coincides with what is adopted by \citet{Carlson94} and by the GNU Scientific
Library. This choice allows to keep the argument $n$ always positive in our formulae. To
compute these integrals we recommend the algorithms by \citet{Fukushima13}. This
method appears faster by the factor of a few in comparison
with the \citet{Carlson94} symmetric forms approach, which was selected by
\citet{AbubGost13} and also adopted in the GNU Scientific Library. \citet{Fukushima13} uses
the following ``associated'' forms of elliptic integrals (note that we also changed sign of
$n$ here):
\begin{equation}
B(m) = \int\limits_0^{\frac{\pi}{2}} \frac{\cos^2\theta\, d\theta}{\sqrt{1-m\sin^2\theta}}, \qquad
D(m) = \int\limits_0^{\frac{\pi}{2}} \frac{\sin^2\theta\, d\theta}{\sqrt{1-m\sin^2\theta}}, \qquad
J(n|m) = \int\limits_0^{\frac{\pi}{2}} \frac{\sin^2\theta\, d\theta}{(1+n\sin^2\theta) \sqrt{1-m\sin^2\theta}}.
\end{equation}
This implies:
\begin{equation}
K(m) = B(m)+D(m), \quad E(m) = B(m)+(1-m) D(m), \quad \Pi(n|m) = K(m)-nJ(n|m).
\end{equation}
Additionally, the following identity from \citep{Fukushima13} becomes useful for us:
\begin{equation}
J(n|m) = -\frac{\pi}{2} \frac{1}{\sqrt{n (1+n) (n+m)}} + \frac{1}{n} K(m) - \frac{m}{n^2} J\left(\frac{m}{n}\Big|m\right).
\label{Jid}
\end{equation}
This identity was used to remove an undesired discontinuity near $\delta=r$ that
appeared in the original MAPLE output, which was expressed via $\Pi$ (see intermediary
quantity $\Omega$ in Table~\ref{tab_Ik}).

Some formulae in Table~\ref{tab_Ik} contain an apparent pecularity near $\delta=0$ due
to division by $\delta$. All functions are actually smooth
at $\delta=0$, but the pecularity is associated with the subtraction of close number
like $K(\epsilon)-E(\epsilon)$ for $\epsilon\approx 0$. This leads to an accuracy loss near
$\delta=0$. To get rid of it, we might rewrite the formulae using some
non-standard elliptic functions as a basis, but this is not convenient, so we choose
to consider the case $\delta\approx 0$ separately and provide in
Table~\ref{tab_Ikd0} the relevant Taylor series about $\delta=0$. These series are
more preferred than general formulae, if $\delta<0.05 (1-r^2)$. We give these series
only for elliptic integrals $n=1$ and only for the case of the full phase of a
transit, $r+\delta<1$. The case of a partial occultation is possible with small $\delta$
only if an additional condition $r\approx 1$ is satisfied. In practice it is a
rare condition when simultaneously $\delta$ is small and $r$ is close to unit (a
close-to-ring eclipse). Besides, numerical tests did not reveal significant loss
of precision in this parametric domain.

Note that our results for $M_0$ (which is responsible only for the flux decrease) are in
agreement with \citet{AbubGost13}. Also, substituting $\delta=0$ and $r=1$ in $M_k$ we
obtain values for the whole-disk momenta $M_k^\star$:
\begin{equation}
M_0^\star = \pi - \frac{\pi}{3} \Lambda_l - \frac{\pi}{6} \Lambda_q, \qquad
M_2^\star = \frac{\pi}{4}-\frac{7\pi}{60}\Lambda_l-\frac{\pi}{15}\Lambda_q, \qquad
M_{1,3}^\star = 0.
\end{equation}

In addition to the expressions~(\ref{Mk}) we may also need
to compute partial derivatives of $M_k$ with respect to $\lambda$, which are trivial, or
with respect to the planet coordinates $x,y$ in the projection plane. We can use the
following formulae for this goal:
\begin{equation}
\delta^2=x^2+y^2, \quad
\frac{\partial\delta}{\partial x} = \frac{x}{\delta}, \quad
\frac{\partial\delta}{\partial y} = \frac{y}{\delta}, \quad
\cos n\lambda = T_n\left(\frac{x}{\delta}\right), \quad
\frac{\partial \cos n\lambda}{\partial x} = n\frac{y^2}{\delta^3} U_{n-1}\left(\frac{x}{\delta}\right), \quad
\frac{\partial \cos n\lambda}{\partial y} = -n\frac{xy}{\delta^3} U_{n-1}\left(\frac{x}{\delta}\right).
\label{cosl}
\end{equation}
Here $T_n$ and $U_n$ are Chebyshev polynomials of the first and second kind.

\begin{landscape}

\begin{table*}
\caption{Integrals
$\mathcal I_{nkj}(\delta,r)$ and their derivatives: general formulae.}
\label{tab_Ik}
\begin{tabular}{@{}clll@{}}
\hline
Function $f$ & $f(\delta,r)$ & $\partial f/\partial r$ & $\partial f/\partial\delta$ \\
\hline
$\mathcal I_{000}$ & $\psi + r^2\varphi - \frac{Q}{2}$ & $2r\varphi$ & $-\frac{Q}{\delta}$ \\
$\mathcal I_{010}$ & $r^2\delta\varphi + (1-\delta^2-r^2) \frac{Q}{4\delta}$ & $2r\delta\varphi-\frac{rQ}{\delta}$ & $r^2\varphi + (r^2-3\delta^2-1)\frac{Q}{4\delta^2}$ \\
$\mathcal I_{020}$ & $r^2\delta^2\frac{\varphi}{2} + \left[3\delta^2(1-r^2-\delta^2)-W\right] \frac{Q}{24\delta^2}$ & $r\delta^2\varphi + (r^2-3\delta^2-1)\frac{rQ}{4\delta^2}$ & $r^2\delta\varphi + \left[W-3\delta^2(r^2+\delta^2+1)\right]\frac{Q}{12\delta^3}$\\
$\mathcal I_{030}$ & $\begin{aligned}[t] r^2\delta^3\frac{\varphi}{4} + &\left[6\delta^2(r^4-3r^2\delta^2-2r^2-\delta^2+1) - \right. \\ &\left. - (1-r^2-3\delta^2)W \right] \frac{Q}{96\delta^3} \end{aligned}$ & $r\delta^3\frac{\varphi}{2}+\left[2W+3\delta^2(r^2-3\delta^2-3)\right]\frac{rQ}{24\delta^3}$ & $\begin{aligned}[t] r^2\delta^2\frac{3\varphi}{4} + &\left[2\delta^2(3r^4-9r^2\delta^2-4r^2-5\delta^2+1) + \right. \\ &\left. + (1-r^2+5\delta^2)W \right]\frac{Q}{32\delta^4} \end{aligned}$ \\
$\mathcal I_{021}$ & $\frac{\psi}{4} + r^2(r^2+2\delta^2)\frac{\varphi}{4} - (5r^2+\delta^2+1)\frac{Q}{16}$ & $r(r^2+\delta^2)\varphi-rQ$ & $r^2\delta\varphi-(r^2+\delta^2+1)\frac{Q}{4\delta}$ \\
$\mathcal I_{031}$ & $r^2\delta (r^2+\delta^2) \frac{3 \varphi}{4} + \left[3(r^2+\delta^2)(1-r^2-\delta^2)-2W\right] \frac{Q}{16\delta}$ & $r\delta (2r^2+\delta^2) \frac{3\varphi}{2} - (1+r^2+5\delta^2) \frac{3rQ}{8\delta}$ & $r^2(r^2+3\delta^2) \frac{3\varphi}{4} + (r^4-5\delta^4-20r^2\delta^2+r^2-5\delta^2-2) \frac{Q}{16\delta^2}$ \\
$\mathcal I_{041}$ & $\begin{aligned}[t] r^2\delta^2 (3r^2+2\delta^2)\frac{\varphi}{4} - \left[(r^2-3\delta^2+3)W + \right. \\ \left. + 6\delta^2(r^4+9r^2\delta^2+2\delta^2+r^2-2)\right]\frac{Q}{96\delta^2} \end{aligned}$ & $\begin{aligned}[t] &r\delta^2 (3r^2+\delta^2) \varphi + \\ & + (r^4-17\delta^4-8r^2\delta^2+r^2-5\delta^2-2) \frac{rQ}{12\delta^2} \end{aligned}$ & $\begin{aligned}[t] r^2\delta (3r^2+4\delta^2) \frac{\varphi}{2} + \left[ (r^2+9\delta^2+3) W + \right. \\ \left. + 6\delta^2 (r^4-15r^2\delta^2-5r^2-4\delta^2) \right] \frac{Q}{48\delta^3} \end{aligned}$ \\
$\mathcal I_{051}$ & $\begin{aligned}[t] &r^2 \delta^3 (2r^2+\delta^2)\frac{5\varphi}{16} + \left[ 30\delta^4(2r^2+\delta^2) (1-r^2-\delta^2) - \right. \\ &\left. - (1+5\delta^4+2\delta^2-r^2+r^2\delta^2) 5W - W^2 \right] \frac{Q}{384\delta^3} \end{aligned}$ & $\begin{aligned}[t] &r\delta^3(4r^2+\delta^2)\frac{5\varphi}{8} + \left[ (3+r^2+37\delta^2) W + \right. \\ &\left. + 6\delta^2 (7r^4-17r^2\delta^2-13r^2-14\delta^2+4) \right] \frac{5rQ}{192\delta^3} \end{aligned}$ & $\begin{aligned}[t] r^2\delta^2(6r^2+5\delta^2) \frac{5\varphi}{16} - \left[(r^2\delta^2-9\delta^4+3r^2-24\delta^2-3) 5W - 3W^2 + \right. \\ \left. + 30\delta^2 (r^4\delta^2+21r^2\delta^4-5r^4+14r^2\delta^2+8r^2+7\delta^2-3)\right] \frac{Q}{384\delta^4} \end{aligned}$ \\
$\mathcal I_{042}$ & $\begin{aligned}[t] &\frac{\psi}{8} + r^2 (r^4+6r^2\delta^2+3\delta^4) \frac{\varphi}{8} - \\ & - (10r^4+\delta^4+19r^2\delta^2+4r^2+\delta^2+1) \frac{Q}{48} \end{aligned}$ & $r (r^4+4r^2\delta^2+\delta^4) \frac{3\varphi}{4} - (1+3r^2+3\delta^2) \frac{3rQ}{8}$ & $r^2\delta (r^2+\delta^2) \frac{3\varphi}{2} + \left[W-3(4r^2\delta^2+r^2+\delta^2)\right] \frac{Q}{8\delta}$ \\
$\mathcal I_{052}$ & $\begin{aligned}[t] r^2\delta &(r^4+3r^2\delta^2+\delta^4) \frac{5\varphi}{8} + \left[(3+r^2+\delta^2)W - \right. \\ &\left. - 6(5r^2\delta^4+5r^4\delta^2+3r^2\delta^2+r^2+\delta^2-1)\right]\frac{5Q}{192\delta} \end{aligned}$ & $\begin{aligned}[t] r\delta &(3r^4+6r^2\delta^2+\delta^4) \frac{5\varphi}{4} + \\ & + \left[ W - 3(7r^2\delta^2+3\delta^4+r^2+2\delta^2)\right] \frac{5rQ}{24\delta} \end{aligned}$ & $\begin{aligned}[t] r^2 &(r^4+9r^2\delta^2+5\delta^4) \frac{5\varphi}{8} - \left[(r^2-7\delta^2+3)W +\right. \\ &\left.+ 6(7r^4\delta^2+23r^2\delta^4+5r^2\delta^2+4\delta^4-r^2-\delta^2+1)\right] \frac{5Q}{192\delta^2} \end{aligned}$ \\
$\mathcal I_{100}$ & $\frac{2}{3}\Omega + \frac{2}{9} \left[ (7r^2+\delta^2-4) \tilde E - W \tilde K \right]$ & $4 r \tilde E$& $\frac{2}{3\delta}\left[ (r^2+\delta^2-1) \tilde E - W\tilde K \right]$ \\
$\mathcal I_{110}$ & $\frac{2}{15\delta} \left[(16r^2\delta^2-W) \tilde E + (1-\delta^2-r^2) W \tilde K\right]$ & $\frac{2r}{3\delta}\left[(r^2+7\delta^2-1) \tilde E - W \tilde K \right]$ & $\frac{2}{15\delta^2} \left\{ (r^2-4\delta^2-1) W \tilde K + \left[\delta^2(19r^2+5\delta^2-5)+W\right] \tilde E \right\}$ \\
$\mathcal I_{120}$ & $\begin{aligned}[t] &\frac{1}{105\delta^2} \left\{ -\left[\delta^2 (9r^2+7\delta^2-7) + 2W \right] W \tilde K + \right. \\ & \left. + \left[8r^2\delta^2 (r^2+15\delta^2-1) + (2r^2-5\delta^2-2) W\right] \tilde E \right\} \end{aligned}$ & $\begin{aligned}[t] \frac{2r}{15\delta^2} &\left\{ (r^2-4\delta^2-1) W \tilde K + \right. \\ & \left. + \left[ \delta^2(4r^2+20\delta^2-5)+W\right] \tilde E\right\} \end{aligned}$ & $\begin{aligned}[t] \frac{1}{105\delta^3} & \left\{ \left[4W-\delta^2(17r^2+21\delta^2+14)\right] W \tilde K - \left[ (4r^2+25\delta^2-4) W + \right.\right. \\ &\left.\left. + \delta^2 (16r^4-35\delta^2-320r^2\delta^2-51r^2+35)\right] \tilde E \right\} \end{aligned}$ \\
$\mathcal I_{130}$ & $\begin{aligned}[t] &\frac{1}{1890\delta^3} \left\{ -\left[\delta^2(224r^2\delta^2+63\delta^2-64r^4+127r^2-63) + \right.\right. \\ &\left.\left. + (8-8r^2-35\delta^2)W\right] W \tilde K + \left[16r^2\delta^4(8r^2+72\delta^2-9) + \right.\right. \\ &\left.\left. + \delta^2(29r^2-27\delta^2-36) W + 8W^2 \right] \tilde E \right\} \end{aligned}$ & $\begin{aligned}[t] &\frac{r}{210\delta^3} \left\{\left[ 3\delta^2(5r^2-21\delta^2-14) + 8W \right] W\tilde K - \right. \\ &\left. - \left[ 3\delta^2(104r^4-232r^2\delta^2-175\delta^2-209r^2+105) + \right.\right. \\ &\left.\left.\phantom{-\big[} + (8r^2+281\delta^2-8)W \right] \tilde E \right\} \end{aligned}$ & $\begin{aligned}[t] &\frac{1}{630\delta^4} \left\{ \left[\delta^4(40r^4+1368r^2\delta^2+39r^2-105+105\delta^2) + \right.\right. \\ &\left.\left. + \delta^2 (13r^2-78\delta^2+36) W - 8W^2 \right] \tilde E + \left[ (8-8r^2+70\delta^2)W + \right.\right. \\ &\left.\left. + \delta^2(83r^4-259r^2\delta^2-125r^2-147\delta^2+42) \right] W \tilde K\right\} \end{aligned}$ \\
$\mathcal I_{121}$ & $\begin{aligned}[t] &\frac{2}{15} \Omega - \frac{1}{225} \left[ (39r^2+9\delta^2+1) W \tilde K - \right. \\ & \left. - (129r^4+9\delta^4-68r^2+246r^2\delta^2-8\delta^2-31) \tilde E \right] \end{aligned}$ & $\frac{2r}{3}\left[ (4r^2+4\delta^2-1) \tilde E - W\tilde K\right]$ & $\frac{1}{15\delta} \left[ (48r^2\delta^2+5r^2+5\delta^2-5-3W)\tilde E - (3r^2+3\delta^2+2) W \tilde K \right]$ \\
$\mathcal I_{131}$ & $\begin{aligned}[t] &-\frac{1}{70\delta} \left\{ (32r^2\delta^2+7r^2+7\delta^2-7-3W) W \tilde K + \right. \\ &\left. + \left[16r^2\delta^2(1-8r^2-8\delta^2) + (4+3r^2+3\delta^2)W \right] \tilde E \right\} \end{aligned}$ & $\begin{aligned}[t] \frac{r}{10\delta} &\left[(88r^2\delta^2+40\delta^4+5r^2-5\delta^2-5-3W) \tilde E - \right. \\ &\left. - (3r^2+13\delta^2+2) W\tilde K \right] \end{aligned}$ & $\begin{aligned}[t]&\frac{1}{70\delta^2} \left\{  - (59r^2\delta^2+21\delta^4-7r^2+7\delta^2+7+3W) W \tilde K + \right. \\ &\left. + \left[\delta^2(152r^4+488r^2\delta^2-19r^2+35\delta^2-35) +  (4+3r^2-18\delta^2)W \right]\tilde E\right\} \end{aligned}$ \\
\hline
\multirow{4}{*}{Auxiliary}
& \multicolumn{3}{c}{
$\psi=\Psi(\delta,1,r) \in \left[0,\pi\right], \quad
 \varphi=\Psi(\delta,r,1) \in \left[0,\pi\right], \qquad
 a=1-(\delta-r)^2, \quad
 b=(\delta+r)^2-1, \quad
 W=ab, \qquad
 Q=\mathcal Q[W], \quad
 Q \text{ nonzero only if } r+\delta>1 \text{ and } |r-\delta|<1;$}\\
& \multicolumn{3}{c}{
$\tilde K = K\left(\frac{4r\delta}{a}\right)\big/\sqrt{a}, \qquad
 \tilde E = \sqrt{a}\, E\left(\frac{4r\delta}{a}\right),\qquad \qquad
 \tilde\Pi = \Pi\left(\frac{4r\delta}{(\delta-r)^2},\frac{4r\delta}{a}\right)\big/\sqrt{a}, \qquad
 \tilde J = a^{-3/2} J\left(\frac{(\delta-r)^2}{a},\frac{4r\delta}{a}\right) - \tilde K \qquad \quad
 \mathrm{if}\; r+\delta<1\; (\text{implies } |r-\delta|<1);$}\\
& \multicolumn{3}{c}{
$\tilde K = K\left(\frac{a}{4r\delta}\right)\big/\sqrt{4r\delta}, \qquad
 \tilde E = \sqrt{4r\delta}\, E\left(\frac{a}{4r\delta}\right) - b\tilde K, \qquad
 \tilde\Pi = \Pi\left(\frac{a}{(\delta-r)^2},\frac{a}{4r\delta}\right)\big/\sqrt{4r\delta}, \qquad
 \tilde J = (4r\delta)^{-3/2} J\left(\frac{(\delta-r)^2}{4r\delta},\frac{a}{4r\delta}\right) - \tilde K \qquad
 \mathrm{if}\; r+\delta>1 \text{ and } |r-\delta|<1;$}\\
& \multicolumn{3}{c}{
$\tilde K = \tilde E = \tilde\Pi = 0 \qquad \mathrm{if} \; |r-\delta|>1\; (\text{implies } r+\delta>1); \qquad
\Omega(\delta,r) = \pi \Theta(r-\delta) + \frac{\delta+r}{\delta-r}\tilde\Pi + (r^2-\delta^2) \tilde K =
                    \frac{\pi}{2} +  (\delta^2-r^2) \tilde J \quad (\text{see text for proof}).$}\\
\hline
\end{tabular}
\end{table*}

\begin{table*}
\caption{Integrals
$\mathcal I_{nkj}$ and their derivatives near $\delta=0$.}
\label{tab_Ikd0}
\begin{tabular}{clll}
\hline
function $f$ & $f(\delta,r)$ & $\frac{\partial f}{\partial r}$ & $\frac{\partial f}{\partial\delta}$ \\
\hline
$\mathcal I_{100}$ & $\frac{2\pi}{3}\left[1-(1-r^2)^{3/2}\right] -\frac{\pi r^2\delta^2}{2\sqrt{1-r^2}}\left[1+\frac{4-r^2}{(1-r^2)^2}\frac{\delta^2}{16}\right]+\mathcal O(\delta^6)$ & $2\pi r \sqrt{1-r^2}\left[1+\frac{r^2-2}{(1-r^2)^2}\frac{\delta^2}{4}+\frac{r^4-8r^2-8}{(1-r^2)^4}\frac{\delta^4}{64}\right] +\mathcal O(\delta^6)$ & $-\frac{\pi r^2\delta}{\sqrt{1-r^2}}\left[1+\frac{4-r^2}{(1-r^2)^2}\frac{\delta^2}{8}\right]+\mathcal O(\delta^5)$\\
$\mathcal I_{110}$ & $\pi r^2\delta\sqrt{1-r^2} \left[1+\frac{3r^2-4}{(1-r^2)^2}\frac{\delta^2}{8}-\frac{r^4-4r^2+8}{(1-r^2)^4}\frac{\delta^4}{64}\right]+\mathcal O(\delta^7)$ & $\begin{aligned}[t]\frac{2\pi r\delta}{\sqrt{1-r^2}} &\left[1-\frac{3}{2} r^2 - \frac{3r^4-8r^2+8}{(1-r^2)^2}\frac{\delta^2}{16}-\right.\\&-\left.\frac{r^6-6r^4+24r^2+16}{(1-r^2)^4}\frac{\delta^4}{128}\right]+\mathcal O(\delta^7)\end{aligned}$ & $\begin{aligned}[t] \pi r^2 \sqrt{1-r^2} &\left[ 1 + \frac{3r^2-4}{(1-r^2)^2}\frac{3\delta^2}{8} - \right. \\ &\left. -\frac{r^4-4r^2+8}{(1-r^2)^4}\frac{5\delta^4}{64}\right]+\mathcal O(\delta^6)\end{aligned}$ \\
$\mathcal I_{120}$ & $\frac{\pi r^2\delta^2}{2\sqrt{1-r^2}}\left[1-\frac{5}{4}r^2-\frac{5r^4-12r^2+8}{(1-r^2)^2}\frac{\delta^2}{16}\right]+\mathcal O(\delta^6)$ & $\begin{aligned}[t]\frac{\pi r\delta^2}{(1-r^2)^{3/2}}&\left[1-3r^2+\frac{15}{8}r^4+\right.\\ &\left.+\frac{5r^6-18r^4+24r^2-16}{(1-r^2)^2}\frac{\delta^2}{32}\right]+\mathcal O(\delta^6)\end{aligned}$ & $\frac{\pi r^2\delta}{\sqrt{1-r^2}}\left[1-\frac{5}{4}r^2-\frac{5r^4-12r^2+8}{(1-r^2)^2}\frac{\delta^2}{8}\right]+\mathcal O(\delta^5)$ \\
$\mathcal I_{130}$ & $\begin{aligned}[t]\frac{\pi r^2\delta^3}{4(1-r^2)^{3/2}} &\left[\vphantom{\frac{r^6}{(1-r^2)^2}}1-\frac{5}{2}r^2+\frac{35}{24}r^4+\right.\\&\left.+\frac{35r^6-120r^4+144r^2-64}{(1-r^2)^2}\frac{\delta^2}{128}\right]+\mathcal O(\delta^7) \end{aligned}$ & $\begin{aligned}[t] &\frac{\pi r\delta^3}{2(1-r^2)^{5/2}} \left[\vphantom{\frac{r^8}{(1-r^2)^2}}1-\frac{9}{2}r^2+\frac{45}{8}r^4-\frac{35}{16}r^6-\right.\\&\left.\phantom{ww}-\frac{35r^8-160r^6+288r^4-256r^2+128}{(1-r^2)^2}\frac{\delta^2}{256}\right]+\mathcal O(\delta^7) \end{aligned}$ & $\begin{aligned}[t]\frac{3\pi r^2\delta^2}{4(1-r^2)^{3/2}} &\left[\vphantom{\frac{r^6}{(1-r^2)^2}}1-\frac{5}{2}r^2+\frac{35}{24}r^4+\right.\\&\left.+\frac{35r^6-120r^4+144r^2-64}{(1-r^2)^2}\frac{5\delta^2}{384}\right]+\mathcal O(\delta^6) \end{aligned}$\\
$\mathcal I_{121}$ & $\begin{aligned}[t]\frac{2\pi}{15} &\left[1-\left(1+\frac{3}{2}r^2\right)(1-r^2)^{3/2}\right]+ \\ &+\frac{\pi r^2\delta^2}{2\sqrt{1-r^2}}\left[1-\frac{3}{2}r^2 - \frac{9r^4-22r^2+16}{(1-r^2)^2}\frac{\delta^2}{32}\right]+\mathcal O(\delta^6)\end{aligned}$ & $\begin{aligned}[t] \pi r \sqrt{1-r^2} &\left[r^2+\frac{9r^4-14r^2+4}{(1-r^2)^2}\frac{\delta^2}{4}+\right. \\ &\left.+\frac{9r^6-32r^4+40r^2-32}{(1-r^2)^4}\frac{\delta^4}{64}\right] + \mathcal O(\delta^6)\end{aligned}$ & $\frac{\pi r^2\delta}{\sqrt{1-r^2}}\left[1-\frac{3}{2}r^2 - \frac{9r^4-22r^2+16}{(1-r^2)^2}\frac{\delta^2}{16}\right]+\mathcal O(\delta^5)$\\
$\mathcal I_{131}$ & $\begin{aligned}[t]\frac{3\pi}{4} r^2\delta\sqrt{1-r^2} &\left[r^2 + \frac{15r^4-24r^2+8}{(1-r^2)^2}\frac{\delta^2}{8} +\right.\\&\left.+ \frac{15r^6-52r^4+64r^2-32}{(1-r^2)^4} \frac{\delta^4}{64} \right]+\mathcal O(\delta^7)\end{aligned}$ & $\begin{aligned}[t] \frac{3\pi r\delta}{\sqrt{1-r^2}} &\left[r^2-\frac{5}{4}r^4 - \frac{45r^6-114r^4+88r^2-16}{(1-r^2)^2}\frac{\delta^2}{32} -\right.\\&\left.- \frac{15r^8-68r^6+120r^4-96r^2+64}{(1-r^2)^4} \frac{\delta^4}{256} \right]+\mathcal O(\delta^7)\end{aligned}$ & $\begin{aligned}[t]\frac{3\pi}{4} r^2\sqrt{1-r^2} &\left[r^2 + \frac{15r^4-24r^2+8}{(1-r^2)^2}\frac{3\delta^2}{8} +\right.\\&\left.+ \frac{15r^6-52r^4+64r^2-32}{(1-r^2)^4} \frac{5\delta^4}{64} \right]+\mathcal O(\delta^6)\end{aligned}$ \\
\hline
& \multicolumn{3}{c}{Assuming that $r+\delta<1$ (full phase of a transit) that implies $r,\delta<1$ and $|r-\delta|<1$.}\\
\hline
\end{tabular}
\end{table*}

\end{landscape}

\section{Investigating model applicability by numerical tests}
\label{sec_num}

As far as our model employs power-series decompositions in $\upsilon$, it should
be applicable only if $\upsilon$ is below some limit. However, such a limit is difficult to
predict theoretically. To determine the actual domain of the applicability, we use
numerical simulations.

\begin{figure*}
\includegraphics[width=0.99\linewidth]{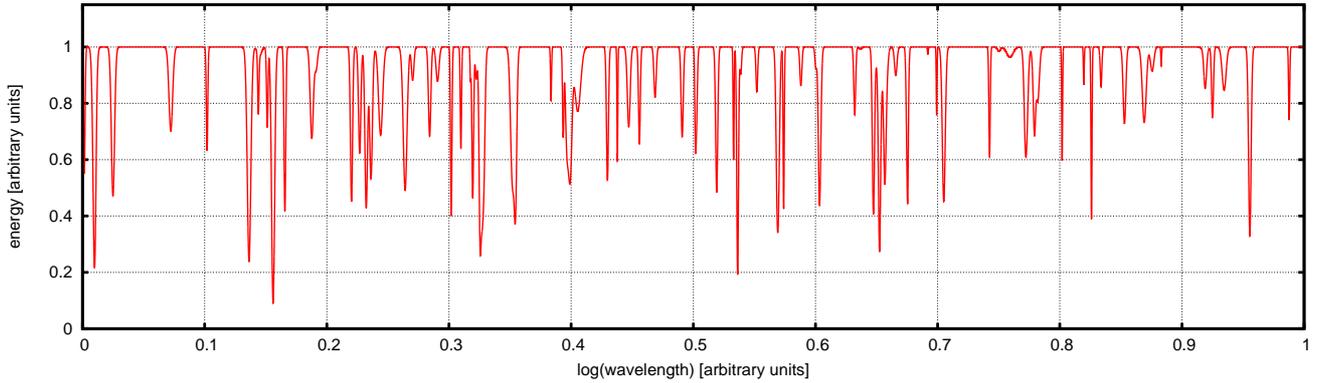}\\
\caption{Synthetic star surface spectrum used in the test simulation (Sect.~\ref{sec_num}).
The spectrum contains $100$ Gaussian lines with random characteristics.}
\label{fig_ssp}
\end{figure*}

First of all, we construct a simulated spectrum $\mathcal F$ containing only Gaussian lines
with randomly chosen characteristics (Fig.~\ref{fig_ssp}). This spectrum we use to perform
direct numerical integrations in~(\ref{spectra}). We compute the out-of-transit spectrum
$\mathcal F_\star$ on a grid of $\upsilon$, and the in-transit ones $\mathcal F_\mathrm{p}$
on a $4$-dimensional grid of the parameters $\upsilon,r,\delta,\lambda$. After that, we fit
numerically each in-transit spectrum $\mathcal F_{\mathrm t}=\mathcal F_\star-\mathcal
F_\mathrm{p}$ with a shifted $\mathcal F_\star$, determining the best fitting shift.
In such a way, we obtain a set of simulated Doppler shifts as a table function
of the gridded values $(\upsilon,r,\delta,\lambda)$. Simultaneously,
we compute our analytic RM model~(\ref{RMtmpl}) on the same grid, and then compare the
results.

However, it is very difficult to seize a four-dimensional space, so we need to
convolve some of the dimensions somehow. We consider $\upsilon$ as our primary parameter of
interest, and for each selected $\upsilon$ we compute only r.m.s. of the
RM model residuals, corresponding to $(r,\delta,\lambda)$ from the grid. The domain of the
grid was constructed as follows: $r\in [0.05,1.22]$,
$\delta\in [\max(r-1,0),1+r]$, $\lambda\in [0,\pi]$. The values of $\upsilon$ were
ranged approximately from $1/3$ to $10$ times the average line width. A subtle but
important detail in this algorithm is that different $r$ imply different amplitudes of the
RM curve, scaling roughly as $r^2$. Therefore, we introduced an additional descaling factor
of $\propto 1/r^2$ to equibalance the contribution of different curves in the cumulative
r.m.s.

To compute the RM model~(\ref{RMtmpl}), we apply three distinct methods to determine the
coefficients $\nu$ and $\mu$. In the first case, we derive them from the original spectrum
$\mathcal F$, which in practice would not be accessible to the observer. In the second
case, we derive $\nu$ and $\mu$ from the observable spectrum $\mathcal F_\star$. And
finally, in the third method we just fit our simulated RM curves
with~(\ref{RMtmpl}), assuming that $\upsilon$, $\nu$, and $\mu$ are all free
regression coefficients. As we discussed above, the first two methods are equivalent for
small $\upsilon$, whenever only the three decomposition terms~(\ref{RMtmpl}) are
significant. But for larger $\upsilon$ more terms enter in the game, making it
important, which of the spectra was in use for $\nu$ and $\mu$. The third method is the one
that can be most easily implemented in practice, if the original spectra are not available
at all.

\begin{figure*}
\includegraphics[width=0.49\linewidth]{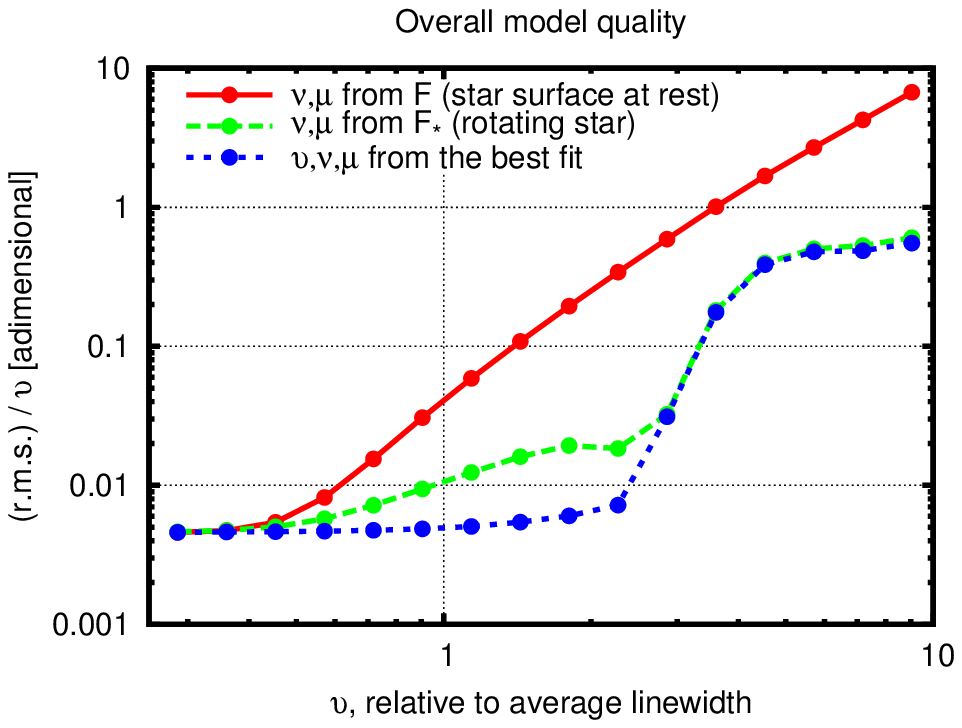}
\includegraphics[width=0.49\linewidth]{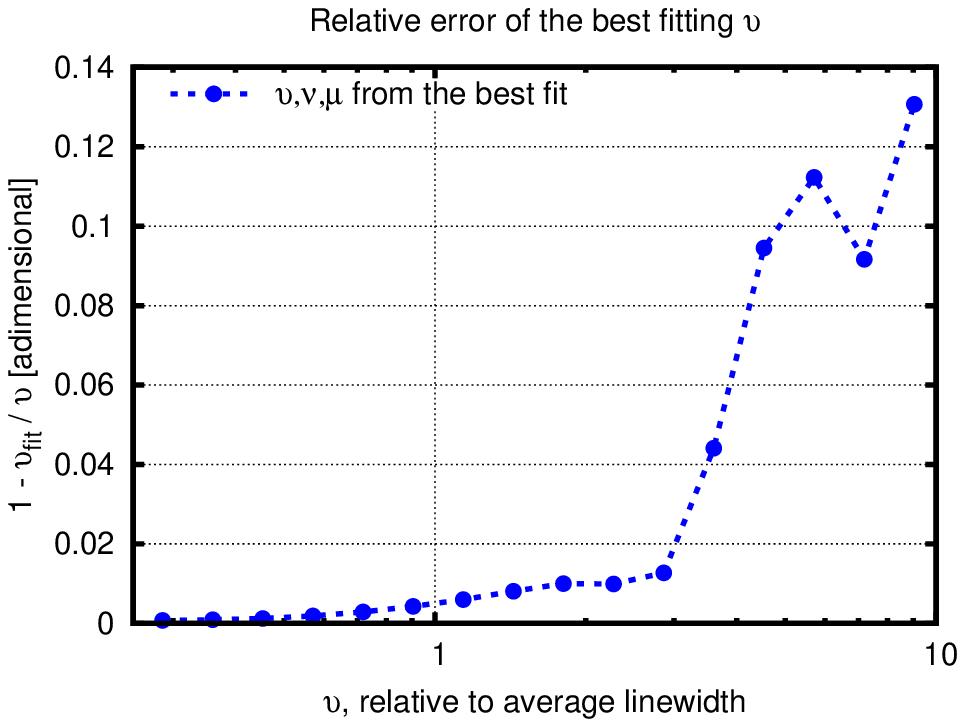}\\
\includegraphics[width=0.49\linewidth]{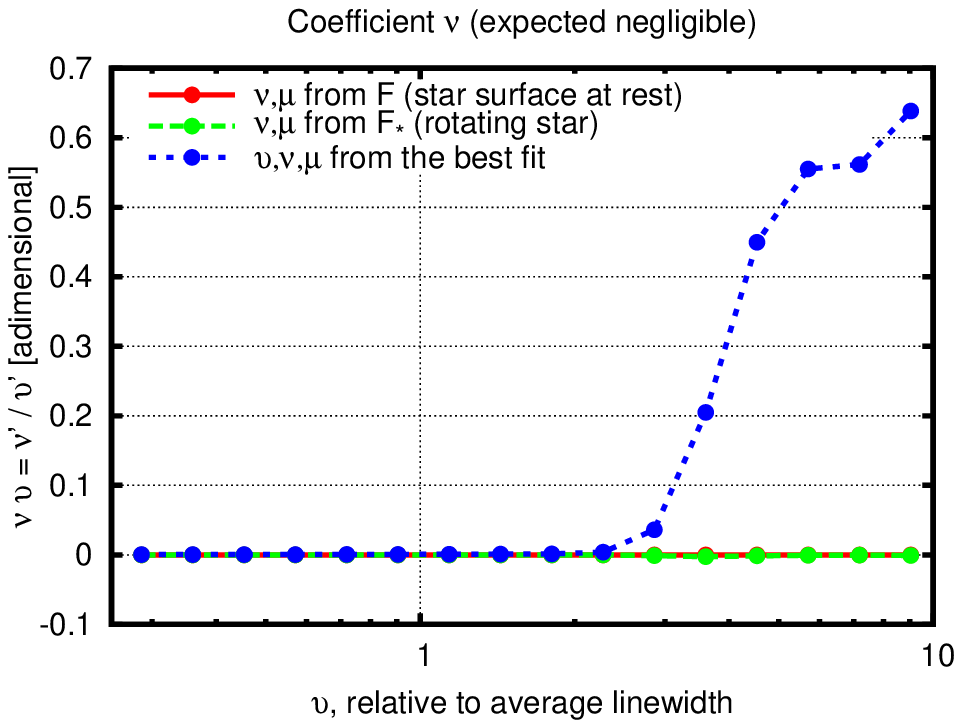}
\includegraphics[width=0.49\linewidth]{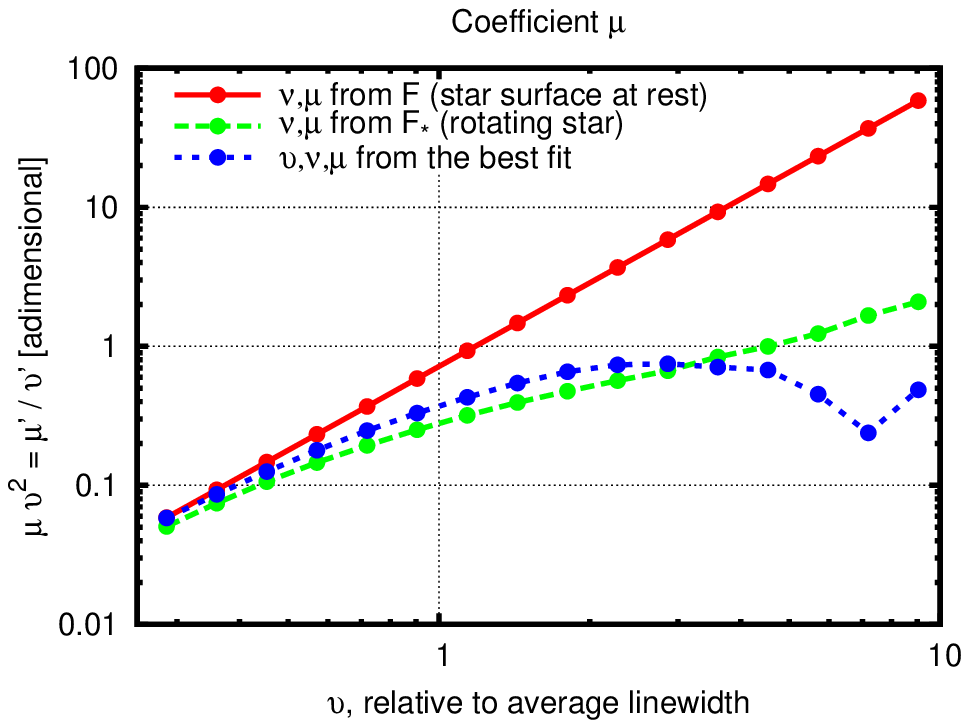}
\caption{Results of the test simulation from Sect.~\ref{sec_num}. The abscissa is the value
of $\upsilon$ normalized by the average linewidth for lines in the original spectrum
$\mathcal F$ (not the rotation-broadened $\mathcal F_\star$). The ordinate
in the first plot is a normalized value (r.m.s.)/$\upsilon$. The value of r.m.s.
also contains an internal normalization by $1/r^2$ (see text), so the graph virtually shows
an average relative error of the model (i.e., residuals are normalized by the amplitudes of
the relevant RM curves). In the last two plots we show adimensional quantities
$\nu\upsilon$ and $\mu\upsilon^2$, which reflect the relative contribution
of the corresponding correction terms in the model~(\ref{RMtmpl}). See text for details and
discussion.}
\label{fig_simul}
\end{figure*}

Our main results of the simulation are demonstrated in Fig.~\ref{fig_simul}. In this
figure, we have four plots that show how the following quantities depend on $\upsilon$: (i)
normalized r.m.s., (ii) relative bias of the
best fitting $\upsilon$, (iii) predicted and best fitting values of $\nu$ (all must be
negligible in our model), (iv) predicted and best fitting values of $\mu$.
From these plots, we can draw the following main conclusions:
\begin{enumerate}
\item It is much better to use $\mathcal F_\star$ for computation of the coefficients $\nu$
and $\mu$, while using $\mathcal F$ infers larger errors. It is very favourable to us,
because the spectrum that we obtain from observations is
also $\mathcal F_\star$, and not $\mathcal F$.

\item If we use $\mathcal F$ to compute $\nu$ and $\mu$, the maximum value of the
rotation velocity, after which our model becomes inacceptable, is only about the average
linewidth (for the lines in the original spectrum $\mathcal F$). If we use $\mathcal
F_\star$ to compute $\nu$ and $\mu$, this limit increases to $2-3$ times the average
linewidth.

\item The best fitting values of $\nu$ and $\mu$ are close to those derived from $\mathcal
F_\star$.
\end{enumerate}

In the summary we may note, that the range of applicability for our model appears
rather optimistic. Even though we used spectra decompositions into powers of $\upsilon$,
our model remains accurate even if $\upsilon$ exceeds the average linewidth. In fact, it is
more adequate to say that our model requires that ``$\upsilon$ is not so large'' instead of
``$\upsilon$ is small''.

\section{Practical application: the case of HD~189733}
\label{sec_appl}

As the number of formulae appearing above was large, let us now describe a concise
step-by-step scheme to compute the RM anomaly:
\begin{enumerate}
\item Compute $\mathcal I_{nkj}$ based on formulae from Table~\ref{tab_Ik}
or~\ref{tab_Ikd0}, if $\delta<0.05(1-r^2)$. Depending on the expected degree of RM anomaly
approximation ($1,2$, or $3$) and degree of the limb-darkening model ($0,1$,
or $2$), not all of these $16$ integrals may be actually needed. Whenever the RM anomaly
curve is not plainly modelled but also fitted based on the RV data, also compute partial
derivatives of $\mathcal I_{nkj}$ to be used in the gradient minimization of the chi-square
or other goodness-of-fit function.
\item Compute momenta $M_k$ from~(\ref{Mk}). Again, not all of them may be
necessary, depending on the particular practical task. If needed for further fitting of the
anomaly, simultaneously compute the derivatives of $M_k$ with respect to
$x,y$, and $r$, based on eq.~(\ref{cosl}).
\item Use formulae~(\ref{RMtmpl}) or~(\ref{RMidn}) to finally compute the RM anomaly. If
necessary, the gradient of the model with respect to the parameters can be
obtained based on~(\ref{Vkdiff}) and on partial derivatives from the previous steps. The
coefficients near $V_{1,2,3}$ can all be treated as free parameters of the fit.
\end{enumerate}

During the transit, the quantities
$\delta$ and $\lambda$ are varying along the projected planet trajectory. This algorithm is
not responsible for modelling the planetary orbital motion during the transit, which
must be carried out separately, e.g. based on a Keplerian or $N$-body model. We omit a
consideration of such models in our paper, as this topic is already investigated quite
well.

We do not take into account the effect of finite light speed that may cause a subtle
time delay between the RV variation due to the planetary orbitaly motion and the
RM anomaly. This delay appears because the former RV shift is imprinted when the light
is emitted from the stellar surface, while the latter one appears when the light is blocked
but the transiting planet, which occurs closer to the observer. This delay should be
usually small, e.g. $\sim 10-20$~sec for a typical hot Jupiter. This effect is not
very hard to model, but this falls out of the scope of the present paper, so we neglect it.

\begin{figure*}
\includegraphics[width=0.99\linewidth]{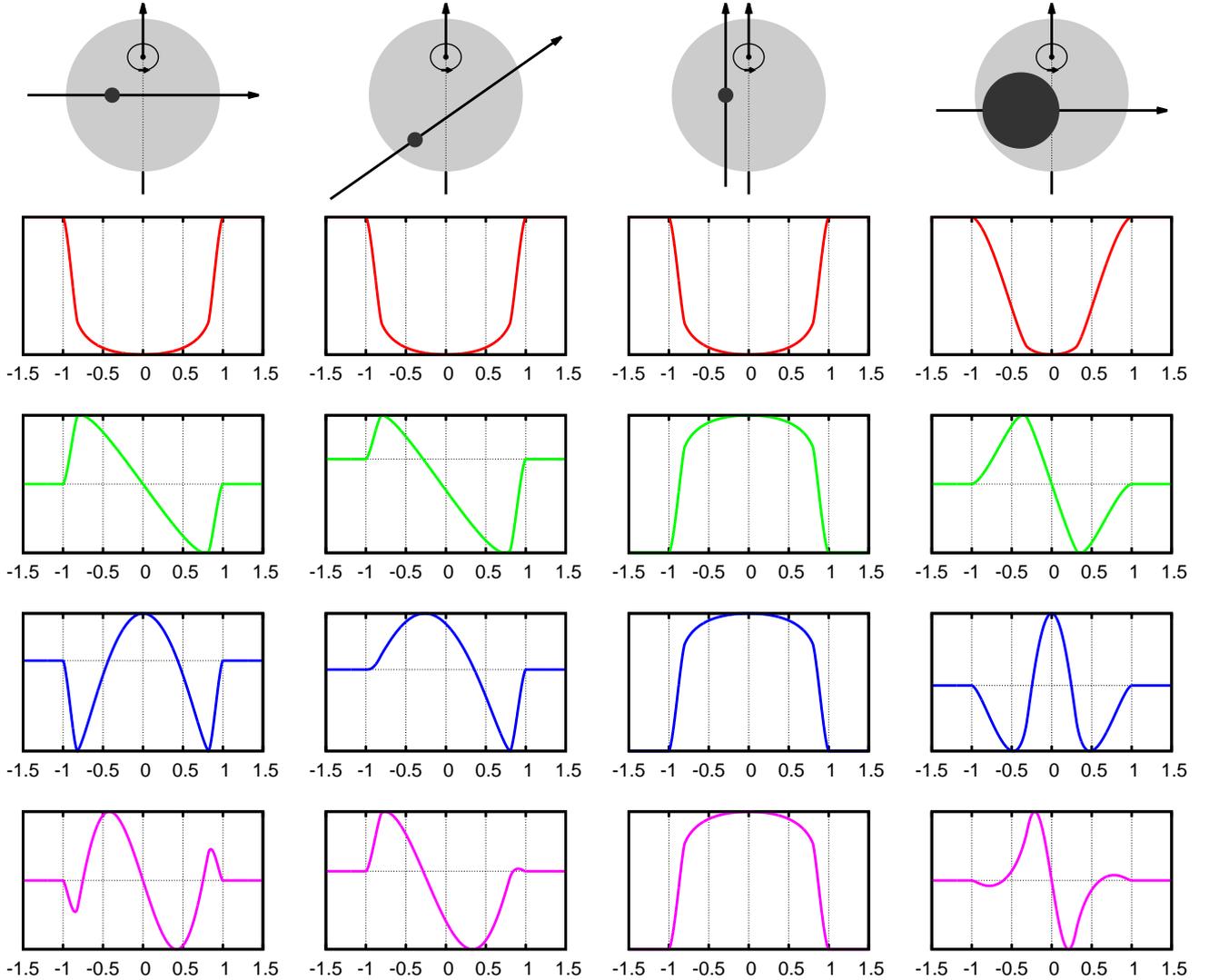}
\caption{Light curves ($1-f$) for several sample transit configurations shown on
top, and plots of the RM anomaly terms $V_1$, $V_2$, and $V_3$ exposed in successive
rows downwards. In each graph, the abscissa is a normalized time with first contact at $-1$
and fourth contact at $+1$. The scale and labels of the ordinates are omitted, because
we intend to demonstrate only shapes of the curves here. The plots assume the limb
darkening with $\Lambda_l=\Lambda_q=0.25$.}
\label{fig_tcrv}
\end{figure*}

First of all, let us gain some impression of the behaviour
of the basis functions $V_{1,2,3}$. During the transit, they can be viewed as functions of
the time, and they also depend on the transit geometry. We plot them in Fig.~\ref{fig_tcrv}
for several sample cases of the planetary orbit orientation. As we can see, the case
in which the planet motion is parallel to the projected star rotation axis, is degenerate.
In this case all $V_{1,2,3}$ have the same or almost the same shape, so it would
be impossible to fit the relevant coefficients separately. But in other cases the shapes of
the basis functions are different, and their coefficients can be fitted independently.

Now we apply our RM effect models to the transiting planet of HD~189733. This is
indended to be just a preliminary and demonstrative study. Full analysis of this and other
objects with the RM effect is prepared for a separate work. HD~189733 is studied very well
already, and it offers an ideally suited a testcase. We use TERRA
\citep{AngladaEscudeButler12} Doppler data derived from the HARPS and SOPHIE
spectroscopy and published in \citep{Baluev15a}. Additionally, we use public Keck RV
data given by \citet{Winn06}, and public transit photometry from
\citep{Bakos06,Winn07a,Pont07}. We do not use a few HARPS-N measurements of this star from
\citep{Baluev15a}, because they appeared entirely erratic after a closer look (this is
being investigated). Also, we do not use vast photometry available for this object
in the Exoplanet Transit Database, as was used in \citep{Baluev15a}. Including
this photometry slows the computations down dramatically without making significant changes
to the models of the RM effect in Doppler data.

We split HARPS data in $5$ independent subsets, corresponding to $4$ transit series and $1$
out-of-transit one. The Keck data were split in two subsets, corresponding
to $1$ in-transit series and to the remaining randomly distributed measurements.
Finally, three Keck points that were obtained before its CCD upgrade in 2004 were removed.
The splitting in such subsets is necessary because the RV scatter on the timescale
of a single in-transit run is only about $\pm 2$~m/s, but on larger transit-to-transit
timescales it increases to $10-20$~m/s. This is likely an activity-induced red noise effect
similar to the one considered e.g. by \citep{Baluev13a}. In the our case of HD~189733 it is
easier to assign fittable RV offsets to different in-transit runs instead of dealing
with correlated noise models as in \citep{Baluev13a}. All Doppler and transit data
were transformed to the same $\text{BJD}_\text{TDB}$ time system using the public software
by \citet{Eastman10}. Note that \citet{Winn06} published their Keck data without performing
the barycentric reduction having amplitude of $\sim 4$~min, and RV data from
\citep{Baluev15a} are in the UTC system, which currently differs from TDB by approximately
1~min. Such differences of a few minutes
become important for self-consistent RV+transits fits.

\begin{figure*}
\includegraphics[width=0.99\linewidth]{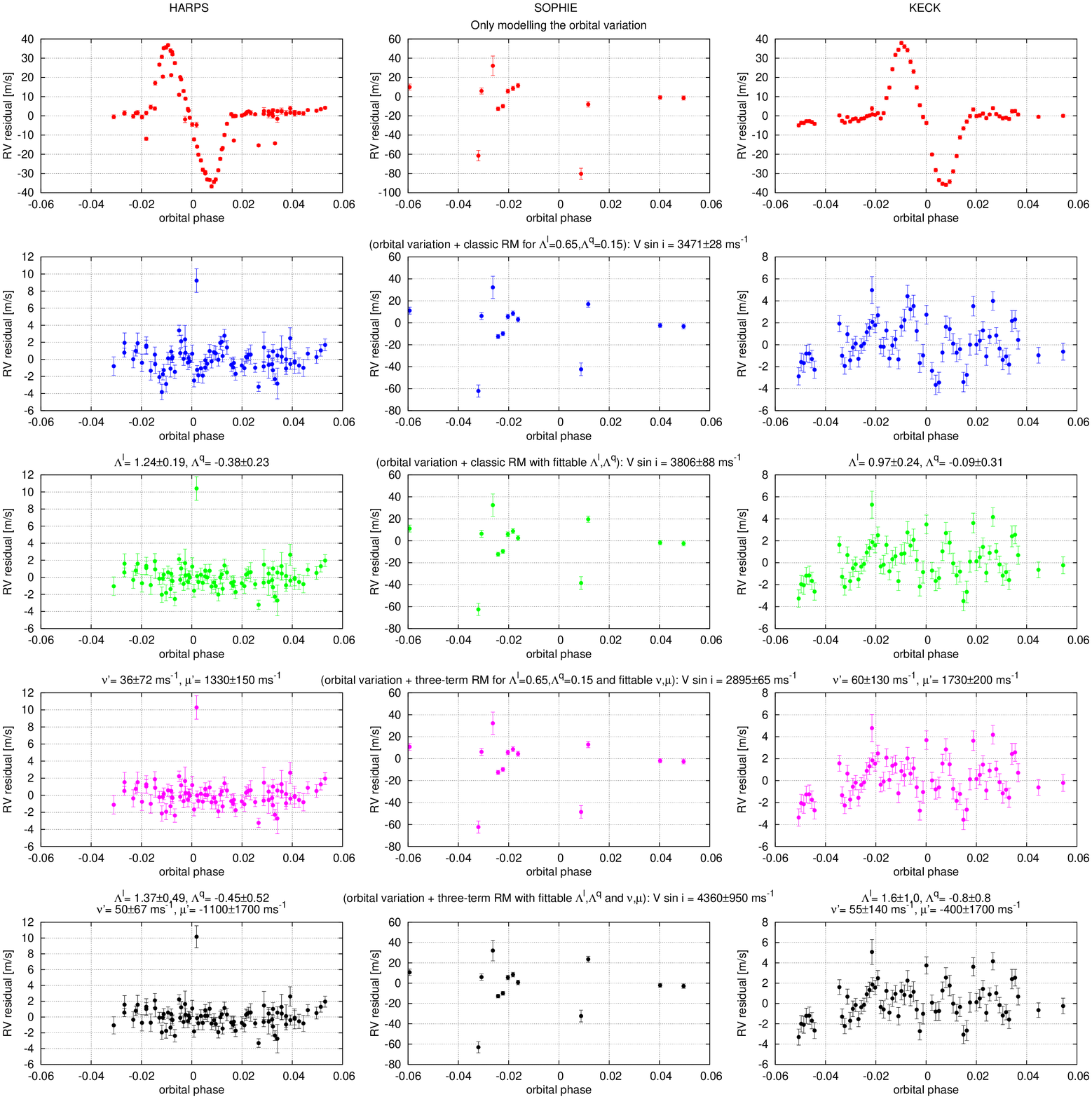}
\caption{Residuals of the Doppler data for HD~189733, computed in the vicinity
of the transit epoch, corresponding to zero abscissa. Each raw of plots corresponds to a
best fitting model labelled above the raw, together with the main best fitting parameters.
See text for a detailed discussion.}
\label{fig_HD189733RM}
\end{figure*}

In Fig.~\ref{fig_HD189733RM} we show RV residuals corresponding to different models, in the
vicinity of the transit, and separately for the HARPS, SOPHIE, and Keck datasets. We can
see that the RM effect is obvious, its curve is well sampled and measured with
high accuracy both by HARPS and Keck, whereas SOPHIE has only sporadic data in
this phase range (top raw of plots).

In the second raw we plot RV residuals of the classic RM model with the limb-darkening
coefficients fixed at $\Lambda^\mathrm{l}=0.65$ and $\Lambda^\mathrm{q}=0.15$. These values
are close to those adopted by \citet{Triaud09}. And confirming \citep{Triaud09}, the
classic model of the RM effect leaves certain systematic wave-like variation
in the residuals, which is clear in HARPS and less clear but noticeable in Keck data. Note
that all $4$ HARPS transit runs are plotted over each other in a single graph. Their
systematic variation cannot be due to the effects like asteroseismologic oscillations,
which would change the shape from one transit to another. This variation
definitely reflects an inaccuracy of the classic RM model.

Formally, this variation can be equally fitted by either (i) adjusting the
limb-darkening coefficients or by (ii) using the correction terms of~(\ref{RMtmpl}). These
ways offer practically equivalent models. The residuals look
almost identical for these fits, and in the both cases they leave no significant hints
of any other systematic variation (third and fourth raws
in Fig.~\ref{fig_HD189733RM}). However, in the case (i), their estimations of the limb
darkening coefficients appear too different from the theoretically predicted values, and
actually do not look physically reasonable. This indicates that the case (ii) is
more realistic. In this case we obtain a well-constrained estimations of the
coefficients $\nu'$ and $\mu'$. The value of $\nu'$ is close to zero, while $\mu'$ appears
comparable to $V\sin i$ (see labels in Fig.~\ref{fig_HD189733RM}). From~(\ref{muG}),
this value of $\mu'$ corresponds to the average width of the spectral lines of $\sim
2/3$ of $V\sin i$, or $\sim 1800$~m/s (before the rotational broadening).

The estimation of $V\sin i$ in the case (ii) is reduced by $\sim 25$ per cent in comparison
with the case (i). This reduced value of $\sim 2900$~m/s is significantly smaller than the
one obtained by \citet{Triaud09} with the classic RM model ($\sim 3300$~m/s) and
even smaller than the value of $\sim 3100$~m/s, obtained by \citet{CCameron10} based on the
line-profile tomography. We need to emphasize that our model is sensitive to the adopted
values of the limb-darkening coefficients, and by increasing of $\Lambda^\mathrm{l}$
we would obtain an estimate of $V\sin i$ closer to \citet{CCameron10}. From the other side,
\citet{CCameron10} use only a linear term in their limb-darkening model, so at the current
stage it is still unclear, which of the two latter estimates is closer to the truth. It is
however definite that all values of $V\sin i$ that rely on the classic
RM model are overestimated.

We also tried to fit simultaneously the RM correction and the limb-darkening
coefficients (fifth raw in Fig.~\ref{fig_HD189733RM}). In this case we obtained an
ill-conditioned fit with large uncertainties, and the residuals did not change. However we
point out that the coefficient $\nu$ is always determined robustly and with a good
accuracy, so it does not seem that there is large correlation between $\nu$ and $\mu$ or
between $\nu$ and the limb-darkening coefficients. Also $\nu$ is always consistent
with zero within narrow limits. This is exactly what
theory predicts, because all Doppler data that we used here are obtained by means of the
spectrum modelling rather than by correlating with a template. In fact, a zero value
of $\nu$ indicates that these spectral models are of a perfect quality.

\section{Conclusions and discussion}

This paper represents an attempt to construct more general but still analytic models of the
RM effect with a particular focus to an improved practical usability, especially
by a third-party analysis work. Although our primary new model~(\ref{RMtmpl}) does not
depend on several important restrictions, like the single-line spectrum, or specific
line profiles, or small planet, there is still much to be done in this topic. The main
vulnerability of this model is that it relies on decompositions in $V\sin i$, requiring it
to be small. In fact, we considered both modelling approaches: employ power-series
decompositions in $V\sin i$, as in \citep{Hirano10}, or avoid such decompositions by
assuming a simple Gaussian line profile, as in \citep{Boue13}. However, our most
useful results correspond to only the first case. In the second case
we did not succeed very much, just showing that the actuality is significantly
more complicated than explicated e.g. by \citet{Boue13}. Nevertheless, we believe that our
primary model can prove a quick and practical workhorse, because most stars that are
involved in planet search programmes are rather quiet, implying that their rotation should
be relatively slow.

Also, we do not consider the effect of macro-turbulence, which was considered e.g.
by \citet{Hirano11,Boue13}, and do not take into account differential rotation of the star.
These and more subtle effects are left for future work.

Regardless of all the remaining limitations, we believe that our model can be useful
in practice, as it is fully analytic, requires nothing but the Doppler data, and can
be applied without detailed knowledge of the spectrum reduction pipelines that depend on a
particular practical case. This paper gives a comprehensive set of all necessary
formula, and we are going to release their implementation with the next version 3.0 of the
{\sc PlanetPack} package \citep{Baluev13c}.

\section*{Acknowledgements}

This work was supported by the Russian Foundation for Basic Research
(project No. 14-02-92615 KO\_a), the UK Royal Society International Exchange grant
IE140055, by the President of Russia grant for young scientists (No. MK-733.2014.2), by the
programme of the Presidium of Russian Academy of Sciences P21, and by the Saint Petersburg
State University research grant 6.37.341.2015. We would like
to thank the anonymous reviewer for useful suggestions and comments on the manuscript.




\bibliographystyle{mnras}
\bibliography{RMac}

\begin{thebibliography}{}
\makeatletter
\relax
\def\mn@urlcharsother{\let\do\@makeother \do\$\do\&\do\#\do\^\do\_\do\%\do\~}
\def\mn@doi{\begingroup\mn@urlcharsother \@ifnextchar [ {\mn@doi@}
  {\mn@doi@[]}}
\def\mn@doi@[#1]#2{\def\@tempa{#1}\ifx\@tempa\@empty \href
  {http://dx.doi.org/#2} {doi:#2}\else \href {http://dx.doi.org/#2} {#1}\fi
  \endgroup}
\def\mn@eprint#1#2{\mn@eprint@#1:#2::\@nil}
\def\mn@eprint@arXiv#1{\href {http://arxiv.org/abs/#1} {{\tt arXiv:#1}}}
\def\mn@eprint@dblp#1{\href {http://dblp.uni-trier.de/rec/bibtex/#1.xml}
  {dblp:#1}}
\def\mn@eprint@#1:#2:#3:#4\@nil{\def\@tempa {#1}\def\@tempb {#2}\def\@tempc
  {#3}\ifx \@tempc \@empty \let \@tempc \@tempb \let \@tempb \@tempa \fi \ifx
  \@tempb \@empty \def\@tempb {arXiv}\fi \@ifundefined
  {mn@eprint@\@tempb}{\@tempb:\@tempc}{\expandafter \expandafter \csname
  mn@eprint@\@tempb\endcsname \expandafter{\@tempc}}}

\bibitem[\protect\citeauthoryear{Abubekerov \& Gostev}{Abubekerov \&
  Gostev}{2013}]{AbubGost13}
Abubekerov M.~K.,  Gostev N.~Y.,  2013, MNRAS, 432, 2216

\bibitem[\protect\citeauthoryear{Anglada-Escud\'e \& Butler}{Anglada-Escud\'e
  \& Butler}{2012}]{AngladaEscudeButler12}
Anglada-Escud\'e G.,  Butler R.~P.,  2012, ApJSS, 200, 15

\bibitem[\protect\citeauthoryear{Bakos et~al.,}{Bakos et~al.}{2006}]{Bakos06}
Bakos G.~A.,  et~al., 2006, ApJ, 650, 1160

\bibitem[\protect\citeauthoryear{Baluev}{Baluev}{2013a}]{Baluev13c}
Baluev R.~V.,  2013a, Astronomy \& Computing, 2, 18

\bibitem[\protect\citeauthoryear{Baluev}{Baluev}{2013b}]{Baluev13a}
Baluev R.~V.,  2013b, MNRAS, 429, 2052

\bibitem[\protect\citeauthoryear{Baluev et~al.,}{Baluev
  et~al.}{2015}]{Baluev15a}
Baluev R.~V.,  et~al., 2015, MNRAS, 450, 3101

\bibitem[\protect\citeauthoryear{Baranne et~al.,}{Baranne
  et~al.}{1996}]{Baranne96}
Baranne A.,  et~al., 1996, A\&ASS, 119, 373

\bibitem[\protect\citeauthoryear{Bou{\'e}, Montalto, Boisse, Oshagh  \&
  Santos}{Bou{\'e} et~al.}{2013}]{Boue13}
Bou{\'e} G.,  Montalto M.,  Boisse I.,  Oshagh M.,   Santos N.~C.,  2013, A\&A,
  550, A53

\bibitem[\protect\citeauthoryear{Butler, Marcy, Williams, McCarthy, Dosanjh  \&
  Vogt}{Butler et~al.}{1996}]{Butler96}
Butler R.~P.,  Marcy G.~W.,  Williams E.,  McCarthy C.,  Dosanjh P.,   Vogt
  S.~S.,  1996, ApJ, 108, 500

\bibitem[\protect\citeauthoryear{Cameron, Bruce, Miller, Triaud  \&
  Queloz}{Cameron et~al.}{2010}]{CCameron10}
Cameron A.~C.,  Bruce V.~A.,  Miller G. R.~M.,  Triaud A. H. M.~J.,   Queloz
  D.,  2010, MNRAS, 403, 151

\bibitem[\protect\citeauthoryear{Carlson}{Carlson}{1994}]{Carlson94}
Carlson B.~C.,  1994, preprint, arXiv.org, math/9409227

\bibitem[\protect\citeauthoryear{Eastman, Siverd  \& Gaudi}{Eastman
  et~al.}{2010}]{Eastman10}
Eastman J.,  Siverd R.,   Gaudi B.~S.,  2010, PASP, 122, 935

\bibitem[\protect\citeauthoryear{Fukushima}{Fukushima}{2013}]{Fukushima13}
Fukushima T.,  2013, J. Comput. \& Applied Math., 253, 142

\bibitem[\protect\citeauthoryear{Gim\'{e}nez}{Gim\'{e}nez}{2006}]{Gimenez06}
Gim\'{e}nez A.,  2006, ApJ, 650, 408

\bibitem[\protect\citeauthoryear{Hirano, Suto, Taruya, Narita, Sato, Johnson
  \& Winn}{Hirano et~al.}{2010}]{Hirano10}
Hirano T.,  Suto Y.,  Taruya A.,  Narita N.,  Sato B.,  Johnson J.~A.,   Winn
  J.~N.,  2010, ApJ, 709, 458

\bibitem[\protect\citeauthoryear{Hirano, Suto, Winn, Taruya, Narita, Albrecht
  \& Sato}{Hirano et~al.}{2011}]{Hirano11}
Hirano T.,  Suto Y.,  Winn J.~N.,  Taruya A.,  Narita N.,  Albrecht S.,   Sato
  B.,  2011, ApJ, 742, 69

\bibitem[\protect\citeauthoryear{Kopal}{Kopal}{1942}]{Kopal42}
Kopal Z.,  1942, Proc. Nat. Acad. Sci., 28, 133

\bibitem[\protect\citeauthoryear{Lanotte et~al.,}{Lanotte
  et~al.}{2014}]{Lanotte14}
Lanotte A.~A.,  et~al., 2014, A\&A, 572, A73

\bibitem[\protect\citeauthoryear{Ohta, Taruya  \& Suto}{Ohta
  et~al.}{2005}]{Ohta05}
Ohta Y.,  Taruya A.,   Suto Y.,  2005, ApJ, 622, 1118

\bibitem[\protect\citeauthoryear{Pepe, Mayor, Galland, Naef, Queloz, Santos,
  Udry  \& Burnet}{Pepe et~al.}{2002}]{Pepe02}
Pepe F.,  Mayor M.,  Galland F.,  Naef D.,  Queloz D.,  Santos N.~C.,  Udry S.,
    Burnet M.,  2002, A\&A, 388, 632

\bibitem[\protect\citeauthoryear{Pont et~al.,}{Pont et~al.}{2007}]{Pont07}
Pont F.,  et~al., 2007, A\&A, 476, 1347

\bibitem[\protect\citeauthoryear{Triaud et~al.,}{Triaud
  et~al.}{2009}]{Triaud09}
Triaud A. H. M.~J.,  et~al., 2009, A\&A, 506, 377

\bibitem[\protect\citeauthoryear{{Welsh} et~al.,}{{Welsh}
  et~al.}{2015}]{Welsh14}
{Welsh} W.~F.,  et~al., 2015, ApJ, 809, 26

\bibitem[\protect\citeauthoryear{Winn et~al.,}{Winn et~al.}{2006}]{Winn06}
Winn J.~N.,  et~al., 2006, ApJ, 653, L69

\bibitem[\protect\citeauthoryear{Winn et~al.,}{Winn et~al.}{2007}]{Winn07a}
Winn J.~N.,  et~al., 2007, AJ, 133, 1828

\makeatother
\end{thebibliography}



\appendix

\section{Rossiter-McLaughlin anomaly for a small planet, arbitrary
rotation velocity, and multi-Gaussian spectra}
\label{sec_rmobs1}

See Sect.~\ref{sec_rmobs} for the details of the approximation method.

\subsection{Cross-correlation with a predefined template}
Let us assume Gaussian approximation for all our spectra:
\begin{equation}
\mathcal F_\star(s) = \mathcal G_{\bm{\beta}_\star}(s,\mathbfit u,\mathbfit c_\star), \qquad
\mathcal F_{\mathrm p}(s) = \mathcal G_{\bm{\beta}_{\mathrm p}(s)}(s-s_{\mathrm p}(s),\mathbfit u,\mathbfit c), \qquad
\mathcal F_\mathrm{T}(s) = \mathcal G_{\bm{\beta}_\mathrm{T}}(s,\mathbfit u_\mathrm{T},\mathbfit c_\mathrm{T}).
\end{equation}
Here the spectral lines positions $\mathbfit u$ are the same for $\mathcal F_\star$ and
$\mathcal F_{\mathrm p}$, but we admit that they may be slightly different from those
used in the template, $\mathbfit u_\mathrm{T}=\mathbfit u + \Delta\mathbfit u$. In this manner we
model possible template imperfections.

Using the expression~(\ref{prodGdiag}) and~(\ref{Fsgauss}), and approximating all slowly
varying functions like $M_0^\star(s)$, and $\sigma_\star(s)$ in the vicinity of each line by
a constant, we can transform equation~(\ref{ccfstar}) to the following:
\begin{equation}
0 = \left\langle \mathcal F_\star \mathcal F_\mathrm{T}' \right\rangle \simeq
\sum_{i=1}^N M_0^\star(u_i) c_{\star,i}c_{\mathrm{T},i} \mathcal G_{\sqrt{\beta_i^2+\sigma_\star^2+\beta_{\mathrm{T},i}^2}}'(-\Delta u_i) =
\frac{1}{\sqrt{2\pi}} \sum_{i=1}^N \frac{M_0^\star(u_i) c_{\star,i}c_{\mathrm{T},i} \Delta u_i}{(\beta_i^2+\sigma_\star^2+\beta_{\mathrm{T},i}^2)^{3/2}} +\mathcal O(\Delta\mathbfit u^3).
\end{equation}
If $\Delta u_i=0$ than this equality is satisfied automatically, and otherwise it sets a
balancing requirement for $\Delta u_i$. For example, it is illegal if all $\Delta u_i$ only
introduce a common systematic shift, because this would just result in a biasing effect on
the RV absolute zero point, which does not affect relative RV measurements that we consider
here.

Various quantities appearing in~(\ref{fitRV}) can be expressed analogously. Dropping the
terms having relative magnitude $\mathcal O(\Delta\mathbfit u^2)$ and smaller, we obtain:
\begin{eqnarray}
\left\langle \mathcal F_{\mathrm p}\mathcal F_\mathrm{T}' \right\rangle &\simeq&
\sum_{i=1}^N M_0(u_i) c_i c_{\mathrm{T},i} \mathcal G_{\sqrt{\beta_i^2+\sigma_{\mathrm p}^2(u_i)+\beta_{\mathrm{T},i}^2}}'(s_{\mathrm p}(u_i))
- \sum_{i=1}^N M_0(u_i) c_i c_{\mathrm{T},i}\Delta u_i \mathcal G_{\sqrt{\beta_i^2+\sigma_{\mathrm p}^2(u_i)+\beta_{\mathrm{T},i}^2}}''(s_{\mathrm p}(u_i)), \nonumber\\
\left\langle \mathcal F_\star \mathcal F_\mathrm{T}'' \right\rangle &\simeq&
- \frac{1}{\sqrt{2\pi}} \sum_{i=1}^N \frac{M_0^\star(u_i) c_{\star,i}c_{\mathrm{T},i}}{(\beta_i^2+\sigma_\star^2(u_i)+\beta_{\mathrm{T},i}^2)^{3/2}},
\qquad \hat s \simeq -\frac{\left\langle \mathcal F_{\mathrm p}\mathcal F_\mathrm{T}' \right\rangle}{\left\langle \mathcal F_\star \mathcal F_\mathrm{T}'' \right\rangle}.
\label{mGs}
\end{eqnarray}

As we can see, the formulae for multiline spectra are significantly more complicated
than for the single-line case considered in previous works. But before discussing them, let
us consider the case of a single line. For $N=1$ the formulae~(\ref{mGs}) reduce to
\begin{equation}
\hat s \simeq - f s_{\mathrm p} \frac{c}{c_\star} \left(\frac{\beta_\star^2+\beta_\mathrm{T}^2}{\beta_{\mathrm p}^2+\beta_\mathrm{T}^2}\right)^{\frac{3}{2}} \exp\left( -\frac{s_{\mathrm p}^2}{2(\beta_{\mathrm p}^2+\beta_\mathrm{T}^2)} \right),
\qquad f=\frac{M_0}{M_0^\star}.
\label{sGs}
\end{equation}
This appears almost equivalent to the formula~(15) by \citep{Boue13}. As they also fit the
template via $\beta_\mathrm{T}$, it becomes equal to $\beta_\star$ in their work.
We obtain an additional factor of $c/c_\star$, the ratio of line intensities in the spectra
of rotating star and stellar surface at rest. This ratio does not appear in \citet{Boue13}.
We believe this factor might be ``lost'' because they put an additional condition that
$\mathcal F_\star$ and $\mathcal F_\mathrm{p}$ both should be pre-normalized, and consider
them containing just a single line without even a continuum. This looks illegal, because
the spectrum normalization mainly depends on its continuum, and not on the
lines. Because line intensity $a_\star$ may be different from $a$, the normalizations of
$\mathcal F_\star$ and $\mathcal F_\mathrm{p}$ become different and cannot not be directly
combined in $\mathcal F_\mathrm{t}$. Instead, it is better to consider unnormalized spectra
treated as energy distributions, as we do in the present work. In this case we still do not
need to take care of the continuum, but spectra normalizations become mutually consistent.

The multiline approximation~(\ref{mGs}) appears even more complicated. First, the summation
over the lines in~(\ref{mGs}) should likely introduce additional broadening effect
in comparison with the single-line formula~(\ref{sGs}). Second, the multiline
expression~(\ref{mGs}) contains terms depending on $\Delta u_i$. This should introduce
additional effect that depends on the quality of the template. This effect was not
characterized previously, because it can be only revealed when working
with the multiline model. Note that the functional shape of this template imperfection
effect should be significantly different from the single-line formula~(\ref{sGs}). Instead
of the dependence on $s_{\mathrm p}$ like $G'(s_{\mathrm p}) \sim s_{\mathrm
p} \exp(-s_{\mathrm p}^2)$ (qualitatively), we should now deal with something like
$G''(s_{\mathrm p}) \sim (s_{\mathrm p}^2-1) \exp(-s_{\mathrm p}^2)$. In fact, we cannot
even guarantee that $\hat s=0$ for $s_{\mathrm p}=0$ in this case:
template imperfections introduce a bias.

Unfortunately, this type of models is very difficult for practical use. It
requires a comprehencive knowledge of deep internals of the spectra processing technique
applied in the particular case. This is not available for authors who want to e.g.
reanalyse some public Doppler data. Moreover, even when such a knowledge is available, the
multiline model becomes mathematically complicated. Therefore, in this work we
do not consider this type of approximations in more details.

We do not give detailed expressions for the case in which the CCF is fitted by a Gaussian,
described by the formula~(\ref{fitRV2f}). Clearly, the final formulae for this case should
be much more complicated than~(\ref{mGs}). Note that due to the template
imperfections, $\Delta \mathbfit u$, appearing for multiline spectra, the
term $\langle\mathcal F_\star \mathcal F_\mathrm{T}'''\rangle$ in~(\ref{fitRV2f}) is
non-zero even for symmetric lines and thus cannot be neglected.

\subsection{Cross-correlation with an out-of-transit stellar spectrum or
parametric modelling of the stellar spectrum (iodine cell technique)}

Now we should just substitute $\mathcal F_\star$ in place of $\mathcal F_\mathrm{T}$ in the
formulae presented above. Formulae~(\ref{mGs}) turn into
\begin{equation}
\left\langle \mathcal F_{\mathrm p}\mathcal F_\star' \right\rangle \simeq
\sum_{i=1}^N M_0(u_i) M_0^\star(u_i) c_i c_{\star,i} \mathcal G_{\sqrt{2\beta_i^2+\sigma_{\mathrm p}^2(u_i)+\sigma_\star^2(u_i)}}'(s_{\mathrm p}(u_i)), \quad
\left\langle \mathcal F_\star \mathcal F_\star'' \right\rangle \simeq
- \frac{1}{4\sqrt\pi} \sum_{i=1}^N \frac{M_0^\star(u_i)^2 c_{\star,i}^2}{(\beta_i^2+\sigma_\star^2(u_i))^{3/2}},
\quad \hat s \simeq -\frac{\left\langle \mathcal F_{\mathrm p}\mathcal F_\star' \right\rangle}{\left\langle \mathcal F_\star \mathcal F_\star'' \right\rangle}.
\end{equation}
Here the template lines misplacements $\Delta u_i$ all vanish, because the new
template coincides with $\mathcal F_\star$. Doppler anomaly can be expressed as follows:
\begin{equation}
\hat s \simeq - \left[\sum_{i=1}^N \frac{M_0(u_i) M_0^\star(u_i) c_i c_{\star,i}}{(\beta_{\mathrm p}^2(u_i)+\beta_\star^2(u_i))^{3/2}} s_{\mathrm p}(u_i) \exp\left(-\frac{s_{\mathrm p}^2(u_i)}{2(\beta_{\mathrm p}^2(u_i)+\beta_\star^2(u_i))}\right)\right]
\left/\left(\sum_{i=1}^N \frac{M_0^\star(u_i)^2 c_{\star,i}^2}{(2\beta_\star^2(u_i))^{3/2}}\right)\right. .
\label{mGs2}
\end{equation}
Now the formula is more simple than~(\ref{mGs}): the template imperfections are irrelevant,
and the RM effect is not biased: $s_{\mathrm p}=0$ implies $\hat s=0$. However, it still
requires a very detailed knowledge of the stellar spectrum.

For a single line, we obtain
\begin{equation}
\hat s \simeq - f s_{\mathrm p} \frac{c}{c_\star} \left(\frac{2\beta_\star^2}{\beta_{\mathrm p}^2+\beta_\star^2}\right)^{\frac{3}{2}} \exp\left( -\frac{s_{\mathrm p}^2}{2(\beta_{\mathrm p}^2+\beta_\star^2)} \right),
\qquad f=\frac{M_0}{M_0^\star}.
\label{sGs2}
\end{equation}
This basically agrees with \citet{Boue13} and \citet{Hirano10}. But again, the
summations in~(\ref{mGs2}) should introduce an additional broadening effect in
comparison with~(\ref{sGs2}).


\bsp	
\label{lastpage}
\end{document}